\documentclass[a4paper,11pt]{article}
\pdfoutput=1
\usepackage[height=8.8in,width=6.45in]{geometry}
\usepackage[font=small,labelfont=bf]{caption}
\usepackage[hidelinks]{hyperref}
\usepackage{multirow}
\usepackage{graphicx}
\usepackage{amssymb, amsmath}
\usepackage{amsfonts}
\usepackage[utf8]{inputenc}
\usepackage{bbold}
\usepackage{bm}
\usepackage{xcolor}
\usepackage{soul}
\usepackage{float}
\usepackage{slashed}
\usepackage{braket}
\usepackage{subcaption}

\newcommand{\ii}{\mathrm{i}}
\newcommand{\rme}{\mathrm{e}}

\newcommand{\vev}[1]{{\left\langle #1 \right\rangle}}

\DeclareMathOperator{\Tr}{Tr}
\DeclareMathOperator{\tr}{tr}
\newcommand{\trp}{\Tr_{\cR}^\prime}

\newcommand{\cusp}{\text{cusp}}

\makeatletter
\newcommand*{\letterdef@}{}
\newcommand*{\letterdef}[3]{%
	\def\letterdef@##1{\expandafter\newcommand\csname #1\endcsname{#2{##1}}}%
	\@tfor\@tempa :=#3\do{\expandafter\letterdef@\expandafter{\@tempa}}}
\makeatother
\letterdef{c#1} {\mathcal}{ABCDEFGHIJKLMNOPQRSTUVWXYZ} 
\letterdef{rm#1}{\mathrm} {dDmM} 

\newdimen\tableauside\tableauside=1.0ex
\newdimen\tableaurule\tableaurule=0.4pt
\newdimen\tableaustep
\def\phantomhrule#1{\hbox{\vbox to0pt{\hrule height\tableaurule
			width#1\vss}}}
\def\phantomvrule#1{\vbox{\hbox to0pt{\vrule width\tableaurule
			height#1\hss}}}
\def\sqr{\vbox{%
		\phantomhrule\tableaustep
		\hbox{\phantomvrule\tableaustep\kern\tableaustep\phantomvrule\tableaustep}%
		\hbox{\vbox{\phantomhrule\tableauside}\kern-\tableaurule}}}
\def\squares#1{\hbox{\count0=#1\noindent\loop\sqr
		\advance\count0 by-1 \ifnum\count0>0\repeat}}
\def\tableau#1{\vcenter{\offinterlineskip
		\tableaustep=\tableauside\advance\tableaustep by-\tableaurule
		\kern\normallineskip\hbox
		{\kern\normallineskip\vbox
			{\gettableau#1 0 }%
			\kern\normallineskip\kern\tableaurule}%
		\kern\normallineskip\kern\tableaurule}}
\def\gettableau#1 {\ifnum#1=0\let\next=\null\else
	\squares{#1}\let\next=\gettableau\fi\next}
\newcommand{\parenth}[1]{\left( #1 \right)}

\renewcommand{\a}{\alpha}
\renewcommand{\b}{\beta}

\newcommand{\pa}{\partial}

\newcommand{\m}{\mu}

\renewcommand{\t}{\tau}

\begin{document}

	\begin{titlepage}
\vbox{
    \halign{#\hfil         \cr
           } 
      }  
\vspace*{15mm}
\begin{center}
{\LARGE \bf 
Emitted Radiation in Superconformal Field Theories
}

\vspace*{15mm}

{\Large Francesco Galvagno}
\vspace*{8mm}

Institut f\"ur Theoretische Physik, ETH Z\"urich
\\
	Wolfgang-Pauli-Strasse 27, 8093 Z\"urich, Switzerland
\vskip 0.8cm
	{\small
		E-mail:
		\texttt{francescogalvagn489@gmail.com}
	}
\vspace*{0.8cm}
\end{center}

\begin{abstract}
The computation of the emitted radiation by an accelerated external particle can be addressed in a gauge theory with the insertion of a Wilson loop. With the addition of conformal symmetry this problem is consistently formalized in terms of correlation functions in presence of the Wilson loop, which are constrained by defect CFT techniques. In theories with extended supersymmetry we can also resort to supersymmetric localization on a four-sphere. By using this set of tools, we review the close relation between the Bremsstrahlung function and the stress energy tensor one-point coefficient in abelian theories and in superconformal field theories. After presenting the state of the art for generic CFTs, we mainly focus on the supersymmetric cases. We discuss the differences between the maximally supersymmetric $\cN=4$ case and $\cN=2$ SCFTs, and finally we review the general and exact result for the emitted radiation in terms of a first order derivative of the Wilson loop expectation value on a squashed sphere.
\end{abstract}

\end{titlepage}

\tableofcontents
\vspace{1cm}

\section{Introduction}
The energy radiated by a particle in an accelerated motion represents an old story within the realm of classical electrodynamics \cite{Dirac:1938nz}, see \cite{Jackson,Landau:1975pou} for a standard textbook approach. 

Such problem is related to a basic question for any Quantum Field Theory (QFT), namely its behavior in presence of an external source. The worldline of the external particle is incorporated at a QFT level as the insertion of an extended operator, the Wilson loop, which represents the phase factor picked up by the accelerating particle. In a pure Maxwell theory, the simplicity of the theory allows one to compute the emitted radiation in an exact way, as a function of the electric charge. The extension to non-abelian gauge theories is more involved, but including additional space-time symmetries (supersymmetry and conformal symmetry) the emitted energy can be understood in terms of special observables in presence of the Wilson loop, and can be computed as a function of the Yang-Mills coupling constant. 

First of all, the emitted radiation is proportional to the so called Bremsstrahlung function $B$, following the relativistic extension of the classical Larmor formula:
\begin{align} \label{DeltaE}
  \Delta E= 2\pi B\! \int \!d\tau \,a^2~,
 \end{align}
where $a$ is the proper acceleration of the particle and $B$ is a theory dependent function of the coupling\footnote{Formula \eqref{DeltaE} presents some subtleties, as discussed at the end of section \ref{sec:2}.}. On the other hand, the Bremsstrahlung function arises naturally in the study of vacuum expectation values of cusped Wilson loops, as a small angle limit of the cusp anomalous dimension \cite{Polyakov:1980ca}\footnote{See section \ref{sec:2} for a proper explanation about cusped Wilson loops.}. 
\begin{equation}\label{Bdef}
\vev{W[C_{\mathrm{cusp}}]} \propto e^{-\Gamma_{\mathrm{cusp}} \log \frac{\Lambda_{\mathrm{UV}}}{\Lambda_{\mathrm{IR}}}}~,~~~~~~~\Gamma_{\mathrm{cusp}} (\varphi) \xrightarrow{\varphi\to 0} -B~\varphi^2~.
\end{equation}
The nontrivial relation between the emitted radiation \eqref{DeltaE} and the Bremsstrahlung function arising from cusp anomalous dimension has been firstly addressed in \cite{Correa:2012at}. In that same paper the Bremsstrahlung function for a probe in the fundamental representation of the gauge group $SU(N)$ has been computed exactly using supersymmetric localization for $\cN=4$ Super Yang-Mills theory. 
Such result was in agreement with previous achievements in AdS/CFT context \cite{Mikhailov:2003er,Athanasiou:2010pv,Hatta:2011gh} and was confirmed by integrability results \cite{Correa:2012hh,Drukker:2012de,Gromov:2012eu,Gromov:2013qga}. Moreover, we consider the interpretation of the Wilson loop as a conformal defect, where a crucial quantity is represented by the displacement operator $\mathbb D_i$, with the index $i$ corresponding to the directions transverse to the defect. This quantity encodes the effects of conformal transformations on correlators in presence of a defect (see also \cite{Billo:2016cpy} for a deeper explanation) and as a defect operator it represents a small perturbation of the line. In \cite{Correa:2012at} it was proven the relation of the Bremsstrahlung function with the two-point function coefficient $C_D$ of the displacement operator:
\begin{equation} \label{displ2point}
\vev{\mathbb D_i(\tau) \mathbb D_j (0)}_W = \frac{C_D\, \delta_{ij}}{\tau^4}~.
\end{equation}
The physical interpretation is that slight modifications of the Wilson loop shape are responsible for the radiation.

Finally, the radiated energy of an accelerated particle can be ``measured'' by computing a flux of the stress tensor of the theory. This idea was originally introduced in \cite{Fiol:2012sg} and fully elaborated \cite{Lewkowycz:2013laa} for $\cN=4$ SYM and led to the conclusion that the energy radiated by the probe can be captured by the insertion of a stress tensor in presence of the Wilson loop. Hence, the Bremsstrahlung function is computed in terms of the one-point coefficient $h_W$ of the stress energy tensor \cite{Kapustin:2005py}:
\begin{equation}\label{hWdef}
\vev{T_{00}(x)}_W= \frac{h_W}{|x_{\perp}|^4}~,
\end{equation}
where $x_{\perp}$ is the orthogonal average distance from the line defect.

Therefore it is reasonable to expect some explicit relations among the observables $\Delta E$, $B$, $C_D$ and $h_W$. 
For abelian gauge theories and in the context of $\cN=4$ SYM there exists an exact correspondence for such observables, and they were computed as exact functions of the coupling constants. 
The approach leading to these results is rather general. Similar results have been obtained for higher representations of the gauge group \cite{Fiol:2011zg,Fiol:2013hna,Fiol:2014vqa}. Also a huge effort has been done in the maximally supersymmetric three-dimensional context, where in a series of papers \cite{Lewkowycz:2013laa,Forini:2012bb,Correa:2014aga,Bianchi:2014ada,Bianchi:2014laa,Bianchi:2017ozk,Bianchi:2017svd,Bianchi:2018bke,Bianchi:2018scb} (see \cite{Drukker:2019bev} for a review) an exact and closed form for the Bremsstrahlung function has been achieved.

The natural question arising at this point is to what extent such results hold when the degree of supersymmetry is decreased. In particular we focus on the $\cN=2$ four-dimensional superconformal case.
The way to compute the emitted radiation in $\cN=2$ theories was conjectured in \cite{Fiol:2015spa}, where the computation of $B$ was associated to a localization computation on a four sphere. Such idea has passed many perturbative checks \cite{Mitev:2014yba,Mitev:2015oty,Gomez:2018usu}, and has been finally proven following two steps: in \cite{Bianchi:2018zpb} the relation between $B$ and $h_W$ was fixed for any $\cN=2$ superconformal theories, whereas in \cite{Bianchi:2019dlw} it was proven that $h_W$ can be computed in terms of the vacuum expectation value of the Wilson loop localized on a squashed sphere. This result is completely general for any superconformal line defect, and therefore it represents a recipe for computing the stress tensor one-point function in presence of an external probe.

The main goal of this paper is to review the computation of the Wilson loop observables related to the emitted energy problem in four dimensional superconformal field theories with extended ($\cN\geq 2$) supersymmetry. After introducing the problem in the free Maxwell case, in section \ref{sec:2} we present the main criticalities for measuring the emitted radiation in a generic CFT. Moving to the superconformal case, in section \ref{sec:3} we review the long series of achievements for $\cN=4$ case, and we also introduce the main techniques that are needed throughout the paper, namely the interpretation of the Wilson loop as a superconformal defect and supersymmetric localization. In section \ref{sec:4} we review the computation of the Bremsstrahlung function in $\cN=2$ conformal theories in terms of a small variation of the background geometry. Finally section \ref{sec:5} is devoted to outline the conclusion and some perspectives, while some technical material is contained in the appendices.

\section{Radiation in free theories}\label{sec:2}
We start by considering the free case, where the simplicity of the theory allows the physical aspects of the problem to emerge naturally.
\subsection{Bremsstrahlung and stress energy tensor in pure Maxwell theory}
\begin{figure}[!t]
	\begin{center}
		\includegraphics[scale=0.7]{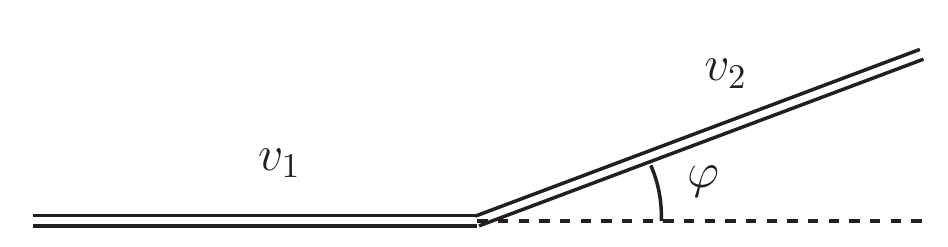}
	\end{center}
	\caption{Cusped Wilson loop with angle $\varphi$.}
\label{fig:Cusp}
\end{figure}

\noindent Let us consider the Maxwell action
\begin{equation}
S_{\mathrm{Max}} = \frac{1}{4} \int d^4x F_{\mu\nu} F^{\mu\nu}~,
\end{equation}
and probe the vacuum with a cusped Wilson line representing the world line of a classical charged particle moving with constant velocity $v_1^\mu$ that suddenly changes to $v_2^\mu$ due to instantaneous acceleration. The parametrization of the line reads:
\begin{equation}
    \begin{aligned}\label{parametrization}
x^\mu&=v_1^\mu\,\tau_1 ~~\mathrm{for}~~ -\infty < \tau_1 < 0 ~, \\
x^\mu&=v_2^\mu\,\tau_2 ~~\mathrm{for}~~ 0 < \tau_2 < +\infty ~.
\end{aligned}
\end{equation}
The velocity vectors satisfy $v_i\cdot v_i=1$ and define the cusp angle $\varphi$ (see Figure \ref{fig:Cusp}) as $ v_1\cdot v_2 = \cos \varphi~$. The cusped Wilson line assumes the following shape:
\begin{equation}\label{WLMaxwell}
W_{\mathrm{cusp}}=\exp\left[\ii\,e\!\int^0_{-\infty}\!\!d\tau_1\,\,v_1^\mu A_\mu(x_1)+
\ii\,e\!\int^{\infty}_0\!\!d\tau_2\,\,v_2^\mu A_\mu(x_2)\right]~,
\end{equation}
where $e$ denotes the electric charge. Since this is a free theory, the only contribution to $\vev{W_{\mathrm{cusp}}}$ is given by the exchange of a single gauge propagator.

\noindent Expanding $W_{\mathrm{cusp}}$ in perturbation theory and using the gauge propagator 
\begin{equation}
    \vev{A_\mu(x_1)A_\nu(x_2)} = \frac{\delta_{\mu\nu}}{4\pi^2x_{12}^2}~, 
\end{equation}
we get \footnote{Following \cite{Grozin:2015kna}, we use dimensional regularization to face the UV divergence and regulate the IR divergence by introducing a dumping factor $\rme^{-\ii\delta(\tau_1-\tau_2)}$ and therefore an IR cut-off $\delta$. In this scheme the cusp anomalous dimension arises as
\begin{equation}\label{Gammadef}
   \vev{W_{\mathrm{cusp}}}=\rme^{-\frac{1}{2\varepsilon}\Gamma_{\mathrm{cusp}}} 
\end{equation}
}:
\begin{align}
    \vev{W_{\mathrm{Cusp}}}=1+e^2 \cos\varphi\,(I(\varphi)-I(0))~,
\end{align}
where:
\begin{align}
I(\varphi) = \!\int\!\frac{d^Dk}{(2\pi)^D}~\frac{1}{k^2\,(k\cdot v_1-\delta)\,(k\cdot v_2-\delta)}~.
\end{align}
This integral has been evaluated explicitly in appendix D of \cite{Bianchi:2019dlw}, so that the final result reads:
\begin{equation}\label{Iphi}
\vev{W_{\mathrm{Cusp}}}=1-\frac{1}{2\varepsilon} \frac{e^2}{4\pi^2}\parenth{\varphi\cot\varphi-1}~.
\end{equation}
Expanding the result for small angles, from \eqref{Bdef} and \eqref{Gammadef} we read the Bremsstrahlung as an exact function in the electric charge:
\begin{equation}\label{Bmaxwell}
    B_{\mathrm{Max}}=\frac{e^2}{12\pi^2}~.
\end{equation}

For this simple theory the relation between the Bremsstrahlung function (and therefore the emitted energy) with the stress tensor one-point function can be verified explicitly. Starting from the Maxwell stress energy tensor, it is possible to evaluate it on the explicit solution for the gauge field $A_\mu$ in presence of a source, given by the Lienard-Wiechert retarded potential, see \cite{Schild:1960,Teitelboim:1979px,Rohrlich} for a thorough analysis. From this solution it is possible to extract the one-point coefficient $h_W$ \cite{Kapustin:2005py}:
\begin{equation}
    h_{W_{\mathrm{Max}}}=\frac{e^2}{32\pi^2}~.
\end{equation}
In summary, two independent computations determine both the Bremsstrahlung function $B$ and the stress tensor coefficient $h_W$ for the abelian Maxwell theory in presence of an external particle. These observables are functions of the coupling, in this case the electric charge $e$, and a simple relation between the two quantities holds:
\begin{equation}\label{Bhmaxwell}
B_{\mathrm{Max}} = \frac{8}{3}h_{W_{\mathrm{Max}}}~.
\end{equation}

\subsection{Physical meaning}\label{sec2.2}
The simplicity of this relation for Maxwell theory is very appealing: the Bremsstrahlung function $B$ parametrizes the radiated energy of the accelerated particle, following \eqref{DeltaE}, while the one-point coefficient $h_W$ defines the energy flux at the infinity, and therefore it represents the ideal observable to capture the emitted energy. A similar simple relation between $B$ and $h_W$ also arise in a different setup, the conformally coupled scalar, where it was found \cite{Fiol:2019woe} that $B=4h_W$. However, already for these simple cases (both Maxwell and the conformal scalar are free theories) we see that the coefficient \eqref{Bhmaxwell} between $B$ and $h_W$ is not universal. Additional issues arise when we try to further generalize this idea.

Equation \eqref{DeltaE} is valid only assuming the initial and final accelerations to be equal and asymptotically vanishing (\textit{i.e.} asymptotically constant velocities, like the configuration chosen in Figure \ref{fig:Cusp}), but in general it is not Lorentz invariant (and in particular it does not hold for any trajectories and for any theories).
This subtlety has generated a strong debate in the past, as summarized in \cite{Fulton,Boulware:1979qj}.

It is possible to define a different object, the invariant radiation rate (see Chapter 5 of \cite{Rohrlich}) as $\cR = v_\mu \frac{dp^\mu}{d\tau}$, where $p^\mu$ is the momentum of the probe particle.\\
This power rate is manifestly Lorentz invariant, since it is not integrated along the worldline, so it properly generalizes the Larmor formula at a relativistic level and can be considered as the correct quantity to measure the emitted radiation of the charged particle. In \cite{Fiol:2019woe} it was found that $\cR$ is directly related to the stress tensor coefficient $h_W$:
\begin{equation}\label{Radrate}
\cR = -\frac{16\pi}{3}\, h_W \, a^\mu a_\mu~.
\end{equation}
However, the subtlety in the relation between $B$ and the total emitted radiation makes the relation between $B$ and $h_W$ rather unclear in general, since the invariant radiation rate does not measure the pure emitted energy due to the presence of non trivial boundary terms\footnote{As mentioned in \cite{Lewkowycz:2013laa}, this fact may be related to the difficulty of separating the radiation component from the self-energy of the particle. \\
Technically, the crucial question should be how to determine the angular distribution of radiation, rather than simply the radiated power. This problem is associated to the explicit shape of the one-point function $\vev{T_{\mu\nu}}_W$ in presence of a time-like defect with an arbitrary worldline. We thank B. Fiol for this private comment.}. Thus we expect no universal relation between $B$ and $h_W$, and in general they will define two different functions of the coupling.

We now want to explore the generalization to non-abelian Yang-Mills theories, where the situation is more involved due to two main issues. As we explained above, it is hard to relate $B$ to $h_W$, also because for Yang-Mills theories conformal invariance is broken at the quantum level. Furthermore, the explicit calculations of $B$ and $h_W$ as functions of the couplings become very complicated in general using standard field theory techniques. \\
Hence, we face these problems in theories that preserve superconformal invariance, and we concentrate on theories endowed with extended supersymmetry with $\cN=4$ and $\cN=2$. We will see that the additional constraints coming from supersymmetry uniquely fix $B$ in terms of $h_W$ and allow the explicit computation of the emitted radiation in terms of the Yang-Mills coupling.

\section{Emitted radiation in $\cN=4$ SYM}\label{sec:3}
We first consider four-dimensional theory endowed with the maximal amount of supersymmetry, namely $\cN=4$ Super Yang-Mills. In this framework, the high degree of symmetry allows one to approach field theory computations with several powerful tools.

\subsection{Wilson loop as a superconformal defect}\label{sec:3.1}
The insertion of a Wilson loop in a superconformal theory generates a partial breaking of the space-time symmetries, as it can be seen as a superconformal defect of the theory. We briefly discuss this symmetry breaking pattern, in order to understand how to constrain observables in this set up.

\paragraph{Broken conformal symmetry}
Throughout the paper we are going to consider a straight line as a conformal defect, so that the four-dimensional euclidean conformal group is broken as
\begin{equation}
\label{sbp}
\mathrm{SO}(1,5)\to \mathrm{SO}(1,2)\times \mathrm{SO}(3)~,
\end{equation} 
which corresponds to 1-d conformal symmetry $\times$ rotations around the line. Such residual symmetry is sufficient to constrain some correlation functions, see \cite{Billo:2016cpy} for a detailed analysis. The two examples that we need are the one-point function of bulk operators  and the two-point function of operators inserted on the defect. Both these observables are fixed in terms of a single coefficient:
\begin{align}\label{1pt2pt}
    \vev{O_\Delta(x)}_W\equiv \frac{\vev{O_\Delta W}}{\vev{W}}= \frac{A_O}{|x_\perp|^\Delta}~,~~~~~\vev{\Phi_{\Delta}(\tau_1)\Phi_{\Delta}(\tau_2)}_W = \frac{C_\Phi}{|\tau_{12}|^{2\Delta}}~.
\end{align}
These formulas can be generalized for objects with spins, the key point is that the kinematics is fixed and the residual physical information resides in the coefficients $A_O$ and $C_\Phi$. 

Explicit examples of these coefficients are the stress tensor coefficient $h_W$ and the displacement normalization $C_D$ as we mentioned in the introduction. In defect CFT the relation between the stress tensor and the displacement operator is extremely meaningful, since the displacement arises as a defect-localized delta function in the stress tensor conservation laws:
\begin{equation} \label{stressTconservationLaw}
\partial_{\mu} T^{\mu i} = - \delta_{D}(x)\mathbb D^{i}+\text{total~derivatives}~.
\end{equation}
The broken translational invariance due to the defect induces the displacement operator to appear as a discontinuity of the stress tensor across the defect, as a sign that energy is exchanged between the defect and the bulk. We expect no universal relation between $C_D$ and $h_W$ by pure conformal constraints (see section 5 of \cite{Billo:2016cpy} for more details), but the further constraints from supersymmetry determine a simple relation between the stress tensor one-point coefficient $h_W$ and the displacement two-point coefficient $C_D$. We will explicitly see this relation in both $\cN=4$ and $\cN=2$ contexts.

\paragraph{Supersymmetric Wilson loop} On the other hand $\cN=4$ is the maximally supersymmetric gauge theory. We explore how supersymmetry enters in the game when including Wilson loops to the theory. 
All the fields are packed in a single vector supermultiplet which contains the gauge field $A_\mu$, four Weyl gauginos $\lambda^A$ and six scalars $\phi^u$. Therefore a supersymmetry-preserving Wilson loop contains a non-trivial coupling with the scalars and takes the form:
\begin{equation}
\label{defWLsusy}
W(C)=\frac{1}{N}\tr \: \mathcal{P}
		\exp \left\{g \oint_C d\tau \Big[\ii \,A_{\mu}(x)\,\dot{x}^{\mu}(\tau)
+ |\dot x|\,n^u(\tau)\phi_u(x)\Big] \right\}~,
\end{equation}
where tr stands for the trace over the fundamental representation of the gauge group ($SU(N)$ throughout this paper), $\cP$ is the path ordering, $g$ is the YM coupling, and $n^u$ is a six-dimensional unit vector specifying the direction of the scalars. Different choices of $n^u$ and the parametrization $x^\mu$ of the line correspond to different fractions of supersymmetry preserved by this insertion. The computation of the Bremsstrahlung function $B$ for $\cN=4$ SYM achieved in \cite{Correa:2012at} highly relies on such different choices, which are allowed by the high amount of supersymmetry.

\subsection{Exact Bremsstrahlung function in $\cN=4$}
\subsubsection*{Half-BPS and latitude Wilson loops}
\begin{figure}[!t]
\begin{minipage}[t]{.55\textwidth}
        \centering
        \includegraphics[scale=0.7]{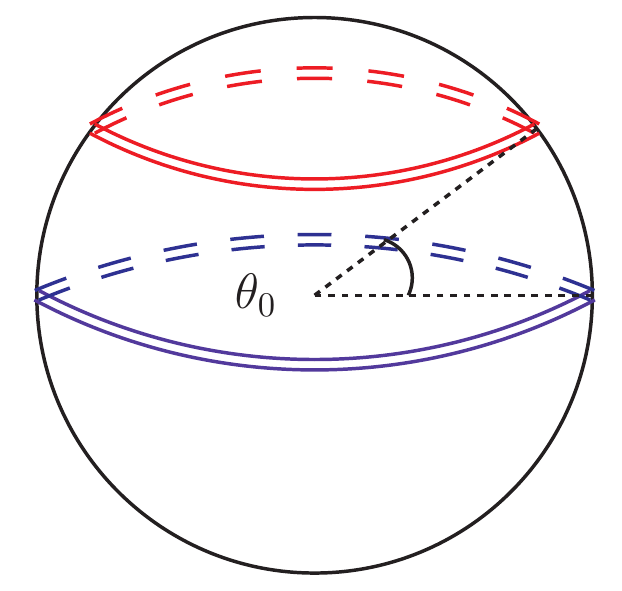}
        \subcaption{1/2-BPS Wilson loop (blue) and 1/4-BPS latitude (red).}\label{Fig:latitude}
    \end{minipage}
    \begin{minipage}[t]{.45\textwidth}
        \centering
        \includegraphics[scale=0.7]{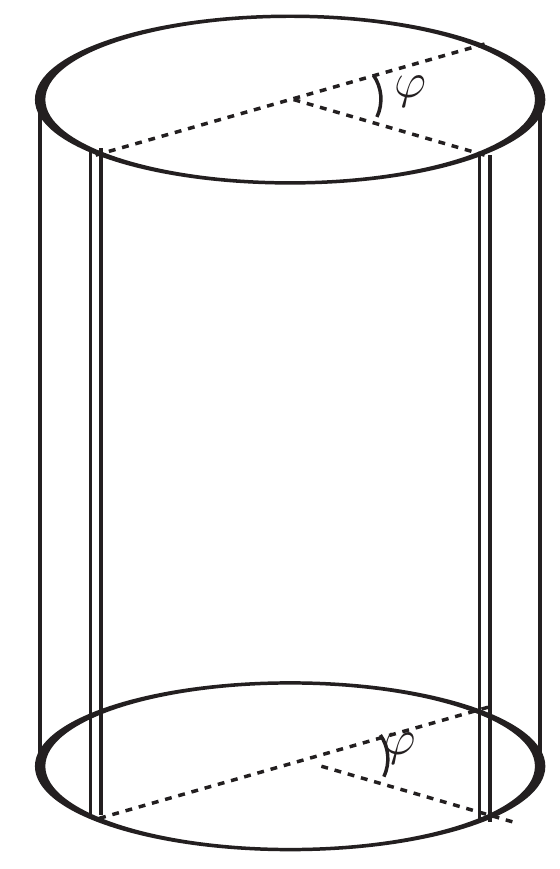} 
        \subcaption{Cusped Wilson loop on the cylinder.}\label{Fig:cuspCylinder}
    \end{minipage}  

    \caption{On the left the $S^4$ configuration, with the half-BPS Wilson loop along the equator and the latitude Wilson loop describing an angle $\theta_0$. \\ On the right the cusped Wilson loop with physical angle $\varphi$ on $\mathbb R\times S^3$.}
\end{figure}

First of all, for a $\tau$-independent $n^{u}$ and considering a circular parametrization of the loop
\begin{equation}\label{circularChoice}
		x^\mu(\tau)=\big(\cos\tau,\sin\tau,0,0\big)~,
\end{equation}
the Wilson loop \eqref{defWLsusy} is maximally supersymmetric (1/2 BPS). This circular Wilson loop has been extensively studied using standard Feynman diagrams \cite{Erickson:2000af,Drukker:2000rr} and has played a central role in the context of supersymmetric localization, which is one of the most special tools of supersymmetric theories. The basic idea is that the path integral of a superconformal theory with extended supersymmetry ($\cN=2,\,4$) only receives contributions from the locus of certain fixed points, and can be reduced to a finite dimensional integral. In appendix \ref{App:mm} we introduce the basics of this procedure and the main results for our purposes, whereas all the technical details can be found in the original paper \cite{Pestun:2007rz} and in the extended review \cite{Pestun:2016zxk}. 
As outlined in appendix \ref{App:mm} the $\cN=4$ partition function can be mapped on a four sphere and expressed as a simple Gaussian integral, see \eqref{ZN4}. The half-BPS circular Wilson loop is mapped on the equator of $S^4$ (see Figure \ref{Fig:latitude}) and its matrix model expression reads:
\begin{equation}
	\label{WLmatrixmodel}
		\mathcal{W}(a)=\frac{1}{N}\,\tr\exp\parenth{\frac{g }{\sqrt{2}}\,a}~,
\end{equation}
so that its expectation value in the Gaussian matrix model of \eqref{ZN4} reads:
\begin{align}
	\label{defWN4}
	\big\langle \cW(a)\big\rangle_{\cN=4} = \frac{1}{\cZ_{\cN=4}} \int da~ \mathrm{e}^{-tr\, a^2} \frac{1}{N} \tr \mathrm{e}^{\frac{g}{\sqrt{2}}\,a} ~.
\end{align}	
This matrix integral can be performed exactly and leads to the exact expression for the Wilson loop vev in the fundamental representation for any values of the coupling $g$ and the number of colors $N$:
\begin{align}
	\label{WN4exact}
		W_{\mathrm{circ}}(g,N) =
		\frac{1}{N}\,L_{N-1}^1\left(\!-\frac{g^2}{4}\right)\,
		\exp\left[\frac{g^2}{8}\Big(1-\frac{1}{N}\Big)\right]~,
\end{align}
where $L_m^n$ are the generalized Laguerre polynomials of degree $m$. This result represents the benchmark for many exact results in $\cN=4$ SYM, and in particular for Wilson loop observables.

A different choice of $n^u(\tau)$ is given by the latitude Wilson loop \cite{Drukker:2006ga,Drukker:2006zk,Drukker:2007dw,Drukker:2007qr,Drukker:2007yx,Pestun:2009nn,Giombi:2009ms,Giombi:2009ds}, parametrized by an angle $\theta_0$ inside the scalar profile:
\begin{equation}
n^u_{\theta_0}=(\sin\theta_0 \cos\tau, \sin\theta_0 \sin\tau, \cos\theta_0,0,0,0)~,
\end{equation}
which defines a latitude on $S^4$, which is 1/4-BPS. We see that for $\theta_0=0$ the 1/2 BPS configuration is restored, see Figure \ref{Fig:latitude}.\\
The vacuum expectation value of the latitude Wilson loop has been computed in \cite{Drukker:2007dw}, in terms the same expression as the circular configuration \eqref{WN4exact}, provided that the coupling constant is redefined in terms of $\theta_0$:
\begin{equation}\label{wthetaWcirc}
\vev{W_{\theta_0}(g)}=\vev{W_{\mathrm{circ}}(\tilde g)}~,~~~~~~~~\tilde g=g\,\cos^2\theta_0~.
\end{equation}

\subsubsection*{Wilson loop variations and scalar two-point functions}
Expanding equation \eqref{wthetaWcirc} around $\theta_0=0$ up to order $\theta_0^2$, and taking $\tilde{g} \sim g\,(1-\theta_0^2)$ one gets:
\begin{equation}\label{vevWthetaexp}
\frac{\vev{W_{\theta_0}}-\vev{W_{\mathrm{circ}}}}{\vev{W_{\mathrm{circ}}}} = -\theta_0^2 \, \frac{1}{2}g\,\partial_g \log \vev{W_{\mathrm{circ}}(g)}~.
\end{equation}
We work out the left hand side of \eqref{vevWthetaexp} by expanding the functional integral in terms of the difference between the unit vectors $n^u_{\theta_0}-n^u$. Again, expanding around small $\theta_0$, for each order in $\theta_0$ we obtain a correlation function of scalars along the line. Since 1-dimensional conformal invariance is preserved along the line defect, the one-point function vanishes and the first non-trivial term reads:
\begin{equation}\label{vevW0vevW}
\frac{\vev{W_{\theta_0}}-\vev{W_{\mathrm{circ}}}}{\vev{W_{\mathrm{circ}}}} = \frac{\theta_{0}^{2}}{2}\int_{0}^{2\pi}\!\! d\tau \int_{0}^{2\pi}\!\! d\tau'\: \hat{n}^{x}(\tau)\,\hat{n}^{y}(\tau')\,\vev{ \phi^x(\tau)\phi^y(\tau')}_{W} +\cO(\theta_0^3)~,
\end{equation}
where $\hat{n}^{x}$ is a 2-dimensional unit vector ($x,y=1,2$). Inside the integrals we recognize a defect two-point function, following the definition:
\begin{equation} \label{f10}
 \vev{\phi^x(\tau)\phi^y(\tau')}_W = \dfrac{\vev{ \tr \bigg[ \cP\, \phi^x(\tau)\, \mathrm{e}^{\int_{\tau}^{\tau'} \left( \ii dx^{\mu} A_{\mu} + |dx|\,  \phi^{u} n_{u}\right)}\phi^y(\tau') \mathrm{e}^{\int_{\tau'}^{\tau} \left( \ii dx^{\mu} A_{\mu} + |dx|\, \phi^{u} n_{u}\right)}\bigg] }}{\vev{W}}~. 
\end{equation}
We evaluate \eqref{vevW0vevW} using the fixed two-point function of \eqref{1pt2pt}. Then imposing the circular parametrization of the loop \eqref{circularChoice} we find:
\begin{equation}
\frac{\vev{W_{\theta_0}}-\vev{W_{\mathrm{circ}}}}{\vev{W_{\mathrm{circ}}}} = \theta_{0}^{2}\, C_\phi\,\frac{\pi}{2}\int_{\epsilon}^{2\pi-\epsilon}\!\!\! d\tau~ \frac{\cos \tau}{1-\cos \tau}  = -\theta_{0}^{2}\,\pi^2\,C_\phi~,
\end{equation}
where $C_\phi$ is the scalar two-point coefficient. We compare this result with \eqref{vevWthetaexp} and obtain:
\begin{equation}\label{CPhi}
C_\phi = \frac{1}{2\pi^{2}}~g\,\partial_{g} \log \vev{W_{\mathrm{circ}}}~,
\end{equation}
which is an exact expression for the coefficient $C_\phi$ in terms of the expectation value of the 1/2-BPS circular Wilson loop \eqref{WN4exact}.

We now consider the Wilson loop configuration which is crucial to extract $B$, namely the cusped Wilson line $W_\text{cusp}$, written as a generalization of \eqref{WLMaxwell} for non-abelian and supersymmetric theories. The parametrization of the line is again given by \eqref{parametrization}, and the velocities define the cusp angle $v_1\cdot v_2 = \cos\varphi$. The explicit expression reads:
\begin{equation}
W_\mathrm{cusp}=\frac{1}{N}\,\tr \cP \exp \,g\,\Big(\int_{-\infty}^0\!d\tau_1\,\mathcal L_1(\tau_1)+
\int_0^{+\infty}\!d\tau_2\,\mathcal L_2(\tau_2)\Big)~,
\label{WcuspN4}
\end{equation}
where $\cL_i(\tau_i)$ are the generalized connections, defined as:
\begin{equation}\label{L1L2Def}
\begin{aligned}	
		\mathcal L_1(\tau_1)&= \ii \,v_1\cdot A(x_1)+\frac{1}{\sqrt{2}}\Big(\mathrm{e}^{+\ii\,\vartheta/2}\,\phi(x_1)
		+\mathrm{e}^{-\ii\,\vartheta/2}\,\bar\phi(x_1)\Big)~,\\
		\mathcal L_2(\tau_2)&= \ii \,v_2\cdot A(x_2)+\frac{1}{\sqrt{2}}\Big(\mathrm{e}^{-\ii\,\vartheta/2}\,\phi(x_2)
		+\mathrm{e}^{+\ii\,\vartheta/2}\,\bar\phi(x_2)\Big)~.
\end{aligned}
\end{equation}
Here $\phi$ is one of the three complex scalar of $\cN=4$ vector multiplet (and $\bar \phi$ is the complex conjugate) and $\vartheta$ is an ``internal'' angular parameter \cite{Drukker:1999zq,Correa:2012nk}, defined as:
\begin{equation}\label{defvartheta}
\cos\vartheta= \vec n_1\cdot \vec n_2~.
\end{equation}
In general the cusped Wilson loop displayed in \eqref{WcuspN4} and \eqref{L1L2Def} is not supersymmetric, but it does become BPS for $\vartheta=\pm \varphi$. Around these values, the Bremsstrahlung function can be extracted from the extended cusp anomalous dimension:
\begin{align}
	\label{GtoBsusy}
		\Gamma_\mathrm{cusp} \simeq
		-\parenth{\varphi^2-\vartheta^2}\,B~,
\end{align}
and $B$ can be obtained either for $\varphi \ll 1$ and $\vartheta=0$ or $\vartheta \ll 1$ and $\varphi=0$. 

Playing with these angles, we can relate the coefficient $C_\phi$ displayed in \eqref{CPhi} with the Bremsstrahlung function.
We switch off the physical angle $\varphi$ and keep the internal angle $\vartheta$. Using a conformal transformation we consider the cylinder configuration, where the cusped Wilson line of Figure \ref{fig:Cusp} is mapped to a quark anti-quark configuration on $\mathbb R \times S^3$, see Figure \ref{Fig:cuspCylinder}. We proceed in the same way as before, by varying $W_\cusp$ around small $\vartheta$, and we obtain a relation between the cusp anomalous dimension with the two-point function of the scalars along the line:
\begin{equation}\label{Gammatheta}
\Gamma_{\mathrm{cusp}}(\varphi=0,\vartheta) = -\frac{\vartheta^{2}}{2} \int_{-\infty}^{\infty}d\tau ~ \vev{ \phi(\tau)\bar\phi(0)}_{W} + \mathcal{O}(\vartheta^{3})~.
\end{equation}
Exploiting again \eqref{1pt2pt} and evaluating the integral, we find that $ \Gamma_\cusp $ for $\varphi=0$ and small $\vartheta$ is written in terms of the the coefficient $C_\phi$:
\begin{equation}\label{GammaCPhi}
\Gamma_{\mathrm{cusp}}(\varphi=0,\vartheta)= C_\phi\, \frac{\vartheta^2}{2} + \cO(\vartheta^{3})~.
\end{equation}
By the comparison of \eqref{GammaCPhi} with \eqref{GtoBsusy} and reading the value of $C_\phi$ from \eqref{CPhi}, we get a formula for the Bremsstrahlung function in terms of the vacuum expectation value of the circular Wilson loop:
\begin{equation}\label{exactBN4}
B(g,N) = \frac{1}{4\pi^{2}}g\,\frac{\partial}{\partial g}\log\vev{ W_{\mathrm{circ}}(g,N)}~.
\end{equation}
Since $\vev{W_{\mathrm{circ}}}$ is an exact expression \eqref{WN4exact}, this result represents an exact formula valid for any values of the coupling constant $g$. As a byproduct, it matches some preliminary weakly and strongly coupled results \cite{Fiol:2011zg,Correa:2012nk,Drukker:2011za}.

\subsection{Bremsstrahlung and displacement operator}
We introduce another crucial observable for the superconformal case, \textit{i.e.} the two-point function of the displacement operator, mentioned in \eqref{displ2point}. The displacement operator $\mathbb D_i$ in a defect CFT is responsible for orthogonal and infinitesimal modifications of the defect. For a time-like Wilson loop the definition is the following:
\begin{align}
    \delta W = \cP\int d\tau \delta x^i(\tau) \mathbb{D}_i(\tau) W~.
\end{align}
Since Wilson lines are special defects explicitly written in terms of fields, we can write an explicit form for the displacement operator:
\begin{align}
    \mathbb{D}_i(\tau) = \ii F_{0j}(\tau) +n^u \partial_j \phi_u(\tau)~.
\end{align}

Using similar tricks as before it is possible to relate the Bremsstrahlung function to the two-point coefficient of the displacement operator (see \cite{Fiol:2013iaa}).
Starting again from the cusped Wilson loop \eqref{WcuspN4} and \eqref{L1L2Def}, we vary $W_\cusp$ with respect to the physical angle $\varphi$. Differently from \eqref{Gammatheta}, $\varphi$ enters in the arguments of both the gauge field and the scalars through the velocities $v_i$. Discarding again $\cO(\varphi)$ terms due to the vanishing of one-point functions along the line, from the variation we precisely get the displacement two-point function:
\begin{equation}\label{gammaCuspDispl}
\Gamma_\cusp \xrightarrow{\varphi\to 0} -\frac{\varphi^2}{2} \int\! d\tau\, \vev{\mathbb{D}_i(\tau) \mathbb{D}_i (0)}_W  + \cO(\varphi^3)~.
\end{equation}
Performing the integral\footnote{Again it is convenient to exploit the conformal map from $S^4$ to $\mathbb R \times S^3$, see Figure \ref{Fig:cuspCylinder}.} and comparing with \eqref{GtoBsusy} we obtain:
\begin{equation}\label{CDequalB}
C_D = 12\, B~.
\end{equation}
Notice that this formula is universal for any four dimensional conformal theory, and in particular its validity can be extended to the $\cN=2$ conformal case, as we will see in the next section.

\subsection{Stress tensor one-point function}
Having introduced the two-point coefficient of the displacement operator, we can relate both $B$ and $C_D$ with the one-point function of the stress tensor, along the line of the discussion of section \ref{sec2.2}. 
In $\cN=4$ case the correspondence between $B$ and $h_W$ comes out from a key observation \cite{Gomis:2008qa,Fiol:2012sg}: the Bremsstrahlung exact formula  \eqref{exactBN4} is analogous to the one-point function of a dimension 2 chiral operator in presence of a Wilson loop. The only difference stands in an overall numerical coefficient. \\
Indeed, defining the chiral operators and their one-point function following \eqref{1pt2pt}:
\begin{equation}
\begin{aligned}\label{vevPhi2}
\Phi_2 (x) =& C_{uv} \tr \phi^u\phi^v(x)~, ~~~C_{uv}~\mathrm{symmetric~ traceless}~,  \\
&~~~~\vev{\Phi_2 (x)}_W = \frac{A_2(g,N)}{4\pi^2|x|^2\phantom{\Big|}}~, 
\end{aligned}
\end{equation}
the computation of $A_2(g,N)$ can be performed using supersymmetric localization on $S^4$. The chiral operators sit in a 1/2-BPS multiplet, and preserve enough supersymmetry to be localized together with the half-BPS Wilson loop: the Wilson loop is placed at the equator and the chiral operator on one of the two poles. The matrix model expression of the chiral operator $\cO(a)$ is slightly subtle due to the non-trivial mixing when moving from $\mathbb R^4$ to $S^4$ (see \cite{Gerchkovitz:2016gxx,Rodriguez-Gomez:2016cem,Rodriguez-Gomez:2016ijh,Billo:2017glv} for further details),
but the resulting matrix integral can be exactly evaluated in terms of Gaussian integrals \cite{Semenoff:2001xp,Pestun:2002mr,Billo:2018oog}:
\begin{equation}
    \begin{aligned}\label{vevO2N4}
 A_2(g,N) &=   \big\langle \cO(a) \cW(a)\big\rangle_{\cN=4} = \frac{1}{\cZ_{\cN=4}} \int da~ \rme^{-tr a^2} \,\cO(a)\, \frac{1}{N}\tr \rme^{\frac{g}{\sqrt{2}}\,a} \\
 &=\frac{1}{8\pi^2}\,g\,\partial_g \log \vev{W_{\mathrm{circ}}}~.
\end{aligned}
\end{equation}

The fact that $B$ and $A_2$ are equal up to a numerical coefficient is clearly not a coincidence. In $\cN=4$ theory the $\Delta=2$ chiral operator belongs to the same supermultiplet as the stress tensor (which is the supercurrent multiplet, see for example \cite{Dolan:2002zh}), so we expect $A_2$ to be related to $h_W$ by a simple numerical coefficient. The exact correspondence between the Bremsstrahlung function $B$ and the stress tensor coefficient $h_W$ in $\cN=4$ SYM has been fixed by \cite{Lewkowycz:2013laa} to be:
\begin{equation}\label{BtohW}
B = 3  \,h_W~.
\end{equation}

\subsubsection*{Summary}
We can summarize the analysis for $\cN=4$. Even for non-abelian theories, but exploiting the presence of maximal supersymmetry, an exact correspondence among the following physical observables holds:
\begin{itemize}
\item
the small angle limit of the cusp anomalous dimension, \textit{i.e.} the Bremsstrahlung function $B$,
\item
the normalization of the displacement operator $C_D$,
\item
the one-point coefficient of the stress tensor $h_W$.
\end{itemize}
Such relations are shown in eqs. \eqref{CDequalB} and \eqref{BtohW} and all of them are related to the total energy emitted by an accelerated heavy particle $\Delta E$ (see \eqref{DeltaE}) as well as the radiation rate $\cR$ (see \eqref{Radrate}). 

The second crucial point is that in $\cN=4$ context the high degree of symmetry is enough to compute all these quantities exactly in terms of the coupling constant $g$: resorting to the exact localization result \eqref{WN4exact} for the 1/2-BPS fundamental Wilson loop vev, $B$, $C_D$ and $h_W$ are computed in terms of the logarithmic derivative of the fundamental Wilson loop vev, see \eqref{exactBN4}. These achievements can also be generalized to any gauge groups and representations \cite{Fiol:2018yuc}.

The next step to address is to what extent these results hold when decreasing the degree of supersymmetry down to $\cN=2$, while keeping conformal invariance. We will see that the $\cN=2$ analysis enriches the physical understanding of the radiation problem.

\section{Radiation and geometry in $\cN=2$ superconformal theories}\label{sec:4}
The problem of the emitted radiation for $\cN=2$ SCFTs has been addressed for the first time in \cite{Fiol:2015spa}, where both the problems of relating $B$ and $h_W$ and finding a consistent way to compute the stress tensor one-point function in a $\cN=2$ SCFT were discussed.

\subsection{Conformal line defects and radiation}

First of all, the connection between the Bremsstrahlung function and the stress tensor coefficient passes through the relation between $B$ and the displacement two-point function. From the $\cN=4$ argument of \cite{Correa:2012at}, we can state that the connection between $B$ and $C_D$ displayed in \eqref{CDequalB} is valid for any conformal theory, since introducing a small angle along the contour is equivalent to the infinitesimal deformations of the line coming from the displacement operator insertions. 

This point has been exploited in \cite{Bianchi:2018zpb} to relate $C_D$ to $h_W$ by only using the constraints of $\cN=2$ superconformal symmetry. Building the whole stress tensor supermultiplet for a generic $\cN=2$ SCFT, the stress tensor conservation law \eqref{stressTconservationLaw} is extended to the other components of the stress tensor multiplet. This procedure generates other defect delta function contributions, corresponding to the components of a supermultiplet of the displacement operator. All the components of the stress tensor multiplet have a one-point function in presence of a Wilson loop which is proportional to $h_W$, and analogously the two-point functions of the whole displacement supermultiplet give rise to the $C_D$ coefficient. Finally superconformal Ward identities provide enough constraints to fix $C_D$ unambiguously in terms of $h_W$: 
\begin{equation}\label{CDtohW}
C_D=36\, h_W~,
\end{equation}
confirming that 
\begin{equation}\label{BtohW2}
    B=3\,h_W~.
\end{equation}

This result is valid for any superconformal line defect with $\cN=2$ supersymmetry. In \cite{Bianchi:2019sxz} a similar computation has determined a similar result $C_D = 48\, h_W$ for a surface defect, proving the complete generality of this procedure.

\subsection{Emitted radiation as a geometry deformation}\label{sec.4.2}
The next step consists in finding the right tool to compute the observables $B,~C_D,~h_W$ in terms of the coupling of the theory, and again we can resort to supersymmetric localization. It turns out that also in $\cN=2$ theories the one-point function of a $\Delta=2$ chiral operator $\vev{\Phi_2 (x)}_W$ can be computed in terms of the localized Wilson loop vev. Indeed equation \eqref{vevO2N4} still holds in $\cN=2$, where the Wilson loop vev is now computed in the $\cN=2$ matrix model, see appendix \ref{App:mm}. However, such one-point function is unrelated to the insertion of the stress-energy tensor, since the $\cN=2$ supercurrent multiplet does not contain any chiral primaries \cite{Dolan:2002zh}. The way to overcome this deadlock has been proposed by \cite{Fiol:2015spa}, following the idea that by definition the stress tensor is associated to a small deformation of the geometry. Therefore, starting from the configuration of a Wilson loop along the equator of a four sphere, the following conjecture has been proposed:
\begin{align}
	\label{conjhW}
	h_W=\frac{1}{12\pi^2}\,\partial_b\log \vev{W_b}\Big|_{b=1} ~.
\end{align}
Here $\vev{W_b}$ is the vacuum expectation value of the Wilson loop on a four-dimensional squashed sphere with squashing parameter $b$ \cite{Hama:2012bg} and for $b=1$ the round sphere is restored. The stress tensor insertion corresponds to a first order deformation of $S^4$, represented by the squashing. This formula has been fully proved in \cite{Bianchi:2019dlw} and represents an optimal way to compute $h_W$, since the right hand side of this relation localizes and can be expressed in terms of a matrix model. We now go through this derivation, which makes use of several techniques from $\cN=2$ SCFTs on curved spaces.

\subsubsection{$\cN=2$ SCFTs on the ellipsoid}
We first define a consistent superconformal theory on the ellipsoid, following \cite{Hama:2012bg}.
We consider a four-dimensional ellipsoid embedded in $\mathbb{R}^5$ as:
\begin{equation}
	\label{defellipsoid2}
		\frac{x_1^2+x_2^2}{\ell^2}+\frac{x_3^2+x_4^2}{\widetilde\ell^{\,2}}+\frac{x_5^2}{r^2}=1~.
\end{equation}
When $\ell=\widetilde\ell=r$ we recover the round sphere. We introduce the squashing parameter $b^2=\ell/\widetilde{\ell}$ so that the axis are parametrized as
\begin{equation}
\label{parametrizationb}
\ell=l(b)\,b~,\quad \widetilde{\ell}=\frac{l(b)}{b}~,\quad
r=r(b)~,
\end{equation}
where the generic functions $l(b)$ and $r(b)$ are such that $l(1)=r(1)=\mathsf r$, $\mathsf r$ being the radius of the sphere. The limit $b\to 1$ recovers the sphere configuration.
We parametrize the ellipsoid in polar coordinates $\xi^\mu=\parenth{\rho,\theta,\varphi,\chi}$:
\begin{align}
	\label{polarcoords}
		x_1&=\ell\, \sin \rho \cos \theta \cos \varphi~,~~~~
		x_2=\ell\, \sin \rho \cos \theta \sin \varphi~,\notag \\
		x_3&=\widetilde{\ell}\,\sin \rho \sin \theta \cos \chi~,~~~~
		x_4=\widetilde{\ell}\, \sin \rho \sin \theta \sin \chi~,~~~~
		x_5=r\, \cos \rho~,
\end{align}
where $\rho\in [0,\pi]$, $\theta\in [0,\pi/2]$, $\varphi\in [0,2\pi]$ and $\chi\in [0,2\pi]$.
From \eqref{polarcoords} we can read the ellipsoid metric $g_{\mu\nu}$, which is also displayed in \eqref{vierbein}.

From \cite{Festuccia:2011ws} it is known that in order to build supersymmetric field theories on a curved space it is necessary to introduce a supergravity multiplet treated as a non-dynamical background, see  \cite{Klare:2013dka} for such a construction on the ellipsoid.
In Euclidean signature, the components of the $\cN=2$ supergravity multiplet are the following \cite{Freedman:2012zz}:
\begin{equation}
	\label{sugra}
		g_{\mu\nu}~,~~
		\psi^{\cI}_\mu~,~~
		\mathsf T_{\mu\nu}~,~~
		\bar{\mathsf T}_{\mu\nu}~,~~
		V_{\mu}^0~,~~
		(V_{\mu})^{\cI}_{\cJ}~,~~
		\eta^{\cI}~,~~
		M~,
\end{equation}
where $g_{\mu\nu}$ is the metric, $\psi^{\cI}_\mu$ (where $\cI=1,2$ R-symmetry index) is the gravitino, 
$\mathsf T_{\mu\nu}$ and $\bar{\mathsf T}_{\mu\nu}$ are real self-dual and anti self-dual tensors\,\footnote{Do not confuse $\mathsf T_{\mu\nu}$, with upright font, with the stress tensor $T_{\mu\nu}$.}, $V_{\mu}^0$ and $(V_{\mu})^{\cI}_{\cJ}$ 
are the gauge fields of the $U(2)_R$ R-symmetry, $\eta^{\cI}$ is the dilatino, and finally $M$ is a scalar field. 
We build a gauge theory consistently coupled to this gravity multiplet:
\begin{align}\label{Sb}
S_b= \frac{1}{g^2}\,\int \!d^4\xi \,\sqrt{\det g}\,L~,
\end{align}
where $L$ contains all the field content for $\cN=2$ theories, hence the gauge vector multiplet $\parenth{A_\mu,\lambda_\cI,\bar \lambda_\cI, \phi, \bar \phi}$ and the matter hypermultiplet $\parenth{q^\cI, \bar q^\cI, \psi, \bar \psi}$. See \cite{Bianchi:2019dlw} for the explicit expression of this Lagrangian.
The ellipsoid action \eqref{Sb} is invariant under $\cN=2$ supersymmetry up to a suitable choice of the non-dynamical supergravity background.
The conditions for having a consistent gauge theory preserving supersymmetry on a curved space are the following:
\begin{equation}\label{susyinv}
\delta_{\mathrm{SUSY}}\, \psi^{\cI}_\mu = 0~, \hspace{1.5cm} \delta_{\mathrm{SUSY}}\, \eta^{\cI}=0~.
\end{equation}
These are defined as the Killing spinor equations, and impose conditions for having  background values for all the bosonic fields $\mathsf T_{\mu\nu},~\bar{\mathsf T}_{\mu\nu},~V_{\mu}^0,~(V_{\mu})^{\cI}_{\cJ},~M$ of the gravity supermultiplet. Their expressions depend on the geometric properties of the ellipsoid, so they are functions of the ellipsoid coordinates $\xi^\m$ and of the squashing parameter $b$. Notice that no background fields should be turned on for a similar construction on the round sphere, therefore the ellipsoid background values shall be considered in their first order variation in the squashing parameter $b$.
The explicit formulas are quite cumbersome, see \cite{Hama:2012bg,Bianchi:2019dlw}, those which are relevant for the present computation can be found in appendix \ref{app:1}.

\subsubsection{Stress tensor and ellipsoid deformation}
We analyze how the vacuum expectation value of the Wilson loop in the superconformal $\cN=2$ theory responds to a small deformation of the ellipsoid geometry, in order to find the relation between the stress tensor coefficient $h_W$ and the 1/2-BPS Wilson loop vev and prove the conjecture \eqref{conjhW}.

\begin{figure}[!t]
	\begin{center}
		\includegraphics[scale=0.7]{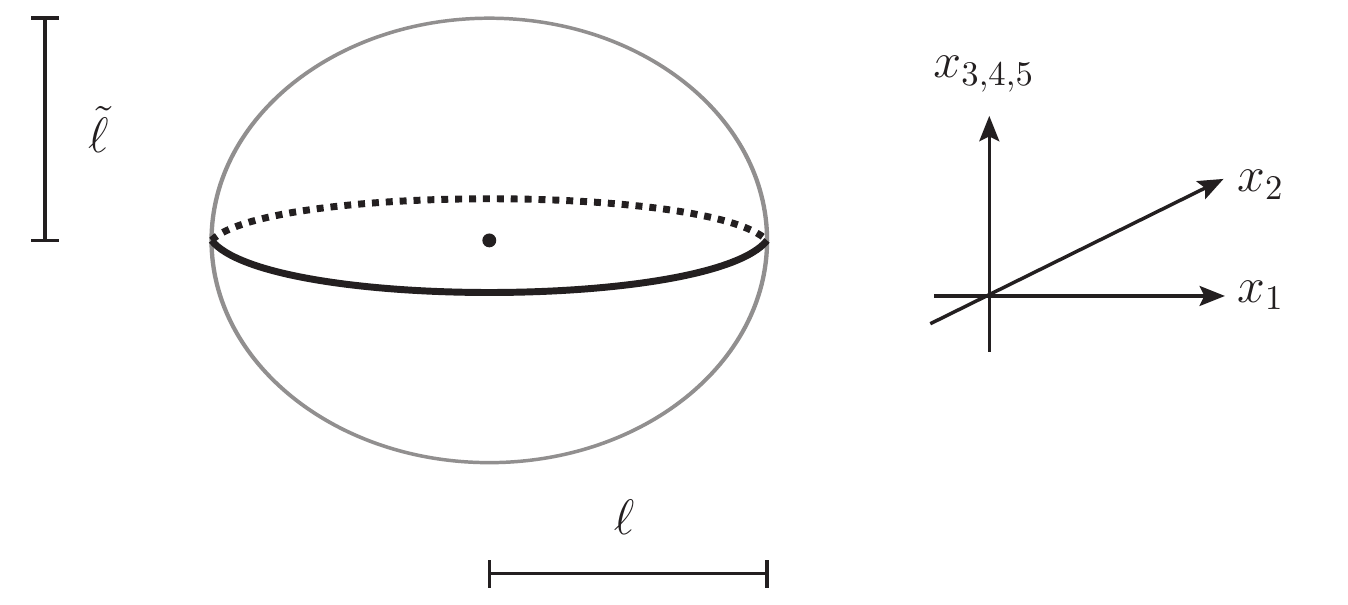}
	\end{center}
	\caption{Wilson loop around the circle of radius $\ell$ in the $x^1,x^2$ plane. In the polar coordinates $\xi^\m$, the
Wilson loop locus $\cC$ is defined by $\chi=\theta=0$, $\rho=\pi/2$.}
	\label{fig:ellipsoid}
\end{figure}

There are two possible half-BPS Wilson loops on the ellipsoid. 
The first configuration wraps the circle of radius $\ell$ in the $x^1,x^2$ plane, the other wraps the circle of radius $\tilde\ell$ in the $x^3,x^4$ plane. 
The two options can be exchanged by sending $b \leftrightarrow b^{-1}$, and are completely equivalent, so we choose the first configuration, see 
Figure~\ref{fig:ellipsoid}.

The explicit expression of the Wilson loop is
\begin{equation}
\label{WLellipsoid}
W_b =\frac{1}{N}\,\tr\,\mathcal{P}
\exp\left[\ii\!\int_\mathcal{C}	\!d\varphi\,\Big(A_\varphi-\ell(\phi+\bar\phi)\Big)\right]~,
\end{equation}
which depends on $b$ through $\ell$, see \eqref{parametrizationb}.
Its vacuum expectation value on the ellipsoid theory is
\begin{equation}\label{vevW}
\big\langle W_b\big\rangle =\frac{1}{Z_b}\,\int \![\cD \Phi] ~ \mathrm{e}^{-S_b}\, W_b~,
\end{equation}
where $S_b$ is defined in \eqref{Sb}.\\
From this definition it follows that
\begin{equation}\label{4_11}
\partial_b\,\log \vev{W_b\!\phantom{\big|}}\Big|_{b=1} =
\frac{-\vev{\partial_b S_b\,W_b\!\phantom{\big|}}
+\vev{\partial_b S_b\!\phantom{\big|}}\,\vev{W_b\!\phantom{\big|}} +\vev{
\!\phantom{\big|}\partial_b W_b}}{\vev{\!\phantom{\big|} W_b}}\bigg|_{b=1}=
-\frac{\vev{\!\phantom{\big|}\! :\!\partial_b S_b\!: W_b}}{\vev{ \!\phantom{\big|}W_b}}\,
\bigg|_{b=1}~.
\end{equation}
A few comments about this expression. Firstly, the dots $:\,:$ indicate the normal ordering, namely the subtraction of all possible self-interactions. Such normal ordering is crucial to ensure the absence of anomalies when computing correlation functions on curved backgrounds, see section 3.2 of \cite{Bianchi:2019dlw} for a detailed discussion on this point. \\
Moreover, in \eqref{4_11} we dropped the term with $\big\langle \partial_b W_b \big\rangle$ since it defines a one-point function along the loop. Due to the preserved conformal symmetry along the defect (see section \ref{sec:3.1}) we expect any defect one-point function to vanish.  Finally let us stress that this expression should not depend on the parametrization \eqref{parametrizationb} of the scales of the ellipsoid.

We work out $\partial_b S_b $, exploiting the fact that the action $S_b$ depends on $b$ only through the background supergravity fields, finding:
\begin{equation}\label{dbSb}
\begin{aligned}
		\partial_b S_b
		= \int \!d^4 \xi\,\sqrt{\det g}&  \bigg[\frac{1}{\sqrt{\det g}}\frac{\partial (\sqrt{\det g}\,L) }{\partial g^{\mu\nu}}\, \pa_b g^{\mu\nu}+
		\frac{\partial L }{\partial (V^{\mu})^{\cJ}{}_{\cI}}
		\,\partial_b (V^{\mu})^{\cJ}{}_{\cI}\\[2mm]
		&+ \frac{\partial L }{\partial \mathsf{T}^{\mu \nu}} \,\partial_b \mathsf{T}^{\mu \nu}
		+  \frac{\partial L }{\pa \bar{\mathsf{T}}^{\mu \nu}} \,
		\partial_b \bar{\mathsf{T}}^{\mu \nu}
		+  \frac{\partial L}{\partial M} \,\partial_b 
		M\bigg]~, 
\end{aligned}
\end{equation}
where the supergravity multiplet was defined in \eqref{sugra}.
By definition, the variation of the action  with respect to the metric at $b=1$
yields the stress-energy tensor $T_{\mu\nu}$ on the sphere. More precisely, we have:
\begin{align}
	\label{defT}
		\frac{\partial (\sqrt{\det g}\,L) }{\partial g^{\mu\nu}}\,\bigg|_{b=1} 
		= - \frac{1}{2} \sqrt{\det g^0}\,\,T_{\mu\nu}~,
\end{align}
where $g^0_{\mu\nu}$ is the metric on the round sphere $S^4$, namely
\begin{equation}
g^0_{\mu\nu}= \lim_{b\to1} g_{\mu\nu}~.
\label{g0sphere}
\end{equation}
Similarly, the variations of the action with respect to the other background fields of the supergravity multiplet 
yield the bosonic components of the \textit{stress-energy tensor supermultiplet}, known also as the supercurrent multiplet \cite{Dolan:2002zh}, given by the following operators:
\begin{align}
    T_{\mu\nu}~,~~J^\cI_\mu~,~~H_{\mu\nu}~,~~\bar H_{\mu\nu}~,~~j_{\mu}~,~~(t_{\mu})_{\mathcal{J}}^{\mathcal{I}}~,~~\chi^\cI~,~~O_2~.
\end{align}
The coefficients relating them to the variations of the Lagrangian must be consistent with the supersymmetry transformations and are crucial for the final result.
We refer to Appendix B of \cite{Bianchi:2019dlw} for the conventions, and the relations read:
\begin{equation}\label{defStressoperators}
\begin{aligned}
&\frac{\partial L }{\partial (V^{\mu})^{\cJ}_{\cI}}\,\Big|_{b=1} 
		= -\frac{\ii}{2} (t_{\mu})_{\mathcal{J}}^{\mathcal{I}}~,~~~~~~~~ \frac{\partial L }{\partial (V^{\mu})^0}\,\Big|_{b=1} 
		= -\frac{\ii}{2} j_{\mu}~,\\[2mm]
\frac{\pa L }{\pa \mathsf{T}^{\mu \nu}}\,\Big|_{b=1} 
		&= -16H_{\mu\nu}~,~~~~~~
\frac{\pa L }{\pa \bar{\mathsf{T}}^{\mu \nu}}\,\Big|_{b=1} 
		= -16\bar H_{\mu\nu}~, ~~~~~~
\frac{\partial L }{\partial M}\,\Big|_{b=1} 
	     = -O_2~.
\end{aligned}
\end{equation}
We substitute these definitions in \eqref{dbSb} and plug the resulting expression for $\partial_b S_b$ in \eqref{4_11} obtaining:
\begin{equation}\label{dlogWvevs}
    \begin{aligned}
\partial_b&\log \big\langle W_b\big\rangle\Big|_{b=1}\! =
\int \!d^4 \xi \,\sqrt{\det g^0} \,
		\bigg[\frac{1}{2}\,\vev{ T_{\mu\nu}}_W \, \pa_b g^{\mu\nu}\big|_{b=1}\!+\!\frac{\ii}{2}\, \vev{
		(t_{\mu})_{\cJ}^{\cI}}_W
		\,\pa_b (V^{\mu})^{\cJ}_{\cI}\big|_{b=1}
		\\[2mm]
		&+\!\frac{\ii}{2}\, \vev{
		j_{\mu}}_W
		\,\partial_b V^{\mu}_0\big|_{b=1}\!+\! 16 \,\vev{ H_{\mu\nu}}_W
		\,\partial_b \mathsf{T}^{\mu \nu}\big|_{b=1}
		\!+\! 16\, \vev{ \bar{H}_{\mu\nu}}_W \,
		\pa_b \bar{\mathsf{T}}^{\mu \nu}\big|_{b=1}\!+\! \vev{ O_2}_W \,\pa_b 
		M\big|_{b=1}\bigg]~.
\end{aligned}
\end{equation}
Here we have dropped the normal ordering symbol $:\,:$ and adopted the short-hand notation $\vev{X}_W$ to identify the normalized one-point function of a normal ordered operator $:\!X\!:$ in presence of the Wilson loop on the sphere. We now want to evaluate such one-point functions and then explicitly calculate the integrals in \eqref{dlogWvevs}. 

\paragraph{Relevant one-point functions:}
The functional form of one-point functions of bulk operators is entirely fixed by the preserved defect 
conformal symmetry, as reviewed in section \ref{sec:3.1}. To write their expressions for the ellipsoid theory, it is convenient to use the so-called embedding formalism (see \cite{Billo:2016cpy}), which realizes conformal transformations as linear transformations over a $d+2$ dimensional space, by exploiting the isomorphism of the conformal group with $SO(1,5)$. Despite some subtleties for operators with spins \cite{Lauria:2018klo}, all the one-point functions in \eqref{dlogWvevs} can be written in terms of the ellipsoid coordinates using this technique.

In presence of a conformal line defect, only operators with even spin can acquire an expectation value \cite{Billo:2016cpy}. Therefore, in
our case, the one-point function of $j_\mu$ and $t_{\mu}$ vanish:
\begin{equation}\label{tmujmu0}
\vev{ (t_{\mu})_{\mathcal{J}}^{\mathcal{I}}}_W=\vev{ j_\mu}_W=0~.
\end{equation}
The only non-zero one-point functions are those of the stress-tensor $T_{\mu\nu}$, of the anti-symmetric tensors $H_{\mu\nu}$ and $\bar{H}_{\mu\nu}$, and the dimension 2 scalar superprimary $O_2$. The explicit formulas for these sphere one-point functions can be found in appendix \ref{app:2}. All these correlators should be generically written in terms of the one-point coefficients, $h_W$, $A_H$ $A_{\bar H}$ and $A_O$ respectively, by using conformal symmetry only.
If we now enforce $\cN=2$ supersymmetry, we can relate all the one-point coefficients to $h_W$ by using superconformal Ward identities. The one-point functions of the whole stress tensor multiplet \eqref{defStressoperators} are written in terms of $h_W$, see \eqref{Tmunu1pt}, \eqref{HbarHsol} and \eqref{O2sol}.

\subsubsection{Integration and final result}
Using these one-point functions together with the the background values of the supergravity multiplet, we have all the necessary ingredients to perform the integration. The explicit results needed to evaluate the sum in \eqref{dlogWvevs} are stored in appendix \ref{app}. 
We begin with the contribution coming from $\vev{T_{\mu\nu}}_W$.
This integral needs to be regularized by a UV cutoff $\epsilon$ to keep the integration away from the defect; the result is
\begin{align}
	\label{resTmunu}
		\int \!d^4 \xi \,\sqrt{\det g^0} ~
		\left[\frac{1}{2}\,\vev{ T_{\mu\nu}}_W \, \partial_b g^{\mu\nu}\big|_{b=1}
		\right]= \parenth{\frac{3l^\prime-3r^\prime
		-3}{\epsilon^3}-\frac{l^\prime-r^\prime-5}{\epsilon}}\,2\pi h_W
		+\cO(\epsilon)~,
\end{align}
where
\begin{equation}
l^\prime=\partial_b l(b)\big|_{b=1}~,\hspace{1cm} r^\prime=\partial_b r(b)\big|_{b=1}~,
\end{equation}
and $l(b)$ and $r(b)$ come from the parametrization \eqref{parametrizationb} of the the ellipsoid axis. The expression \eqref{resTmunu} does not contain any finite contribution, whereas the divergence is an artifact of the regularization procedure. In particular, computing the integral \eqref{resTmunu} in dimensional regularization would simply return a vanishing result. Hence we discard this contribution.\\
Instead, the other terms in \eqref{dlogWvevs} yield finite contributions:
\begin{subequations}
\begin{align}
\int \!d^4 \xi \,\sqrt{\det g^0} ~
		\Big[16 \,\big\langle H_{\mu\nu}\big\rangle_W
		\,\partial_b \mathsf{T}^{\mu \nu}\big|_{b=1} \Big] &=\big(14+4l^{\prime}
		-4r^{\prime}\big)\pi^2 h_W-\frac{3}{2}\pi^4h_W~,\label{resfinH}\\[2mm]
\int \!d^4 \xi \,\sqrt{\det g^0} ~
		\Big[16 \,\big\langle \bar{H}_{\mu\nu}\big\rangle_W
		\,\partial_b \bar{\mathsf{T}}^{\mu \nu}\big|_{b=1} \Big]&=\big(14+4l^{\prime}
		-4r^{\prime}\big)\pi^2 h_W-\frac{3}{2}\pi^4h_W~,\label{resfinHbar}\\[2mm]
\int \!d^4 \xi \,\sqrt{\det g^0} ~
		\Big[\big\langle O_2\big\rangle_W \,\partial_b 
		M\big|_{b=1}\Big]&=-\big(16+8l^{\prime}
		-8r^{\prime}\big)\pi^2 h_W+3\pi^4h_W~.\label{resfinO2}
\end{align}
\end{subequations}
We notice that the individual integrals depend on
$l^\prime$ and $r^\prime$, related to the arbitrary chosen parametrization of the ellipsoid \eqref{parametrizationb}. However, when we sum \eqref{resfinH}, \eqref{resfinHbar} and \eqref{resfinO2} the result is independent of such a choice. Therefore, collecting all the finite contributions we can rewrite \eqref{dlogWvevs} as follows:
\begin{equation}\label{dlogWhW}
\partial_b\log \big\langle W_b\big\rangle\Big|_{b=1} =12\pi^2 h_W~.
\end{equation}
This formula exactly proves the conjecture \eqref{conjhW} and represents an exact way of computing the stress tensor one-point function in presence of an external probe, and due to \eqref{CDtohW} it also represents a way to measure the Bremsstrahlung function of the accelerated particle in $\cN=2$ superconformal settings, according to the following formula:
\begin{equation}\label{BdlogW}
B=\frac{1}{4\pi^2}\,\partial_b\log \big\langle W_b\big\rangle\Big|_{b=1}~.
\end{equation}

\subsection{Ellipsoid matrix model and perturbative results}
The proven formula \eqref{conjhW} between the one-point coefficient $h_W$ and the first order variation of the ellipsoid Wilson loop purely relies on superconformal symmetry of the ellipsoid gauge theory. We did not use any explicit realization of the gauge theory in terms of fundamental fields. We now show that formula \eqref{conjhW} is useful to evaluate the stress tensor coefficient as a function of the coupling. Again, the crucial technique is represented by supersymmetric localization. Indeed, in \cite{Hama:2012bg} SUSY localization was applied to the ellipsoid theory to express its partition function and the expectation value of circular Wilson loops in terms of a matrix model. Hence $h_W$ can be evaluated using matrix model techniques. 
Since this is a $\cN=2$ theory, the computation of the Wilson loop vev in a matrix model is more involved than the $\cN=4$ case described in \eqref{WLmatrixmodel}.
The Wilson loop \eqref{WLellipsoid} on the localization locus keeps the same shape than the $\cN=4$ case, apart from an explicit dependence on the squashing $b$:
\begin{align}
	\label{wlMMellipsoid}
		\cW_b(a) = \frac{1}{N}\tr \exp \Big(\frac{b\,g}{\sqrt{2}}\, a\Big)~.
\end{align}
Instead, the partition function contains the Gaussian term, like the $\cN=4$ case, but also a one-loop determinant and an instanton term, which are specific of $\cN=2$ theories:
\begin{align}\label{Zb}
		\cZ_b = \int \! da~\mathrm{e}^{-\tr a^2} \,\big|\cZ^\text{1-loop}_b\big|^2	\, \big|\cZ^\text{inst}_b\big|^2~.
\end{align}
Luckily, $\cZ_b$ enjoys some special properties that allow to perform the matrix integral.
Both the one-loop determinant and the instanton term only depend on the squashing parameter $b = (\ell/\tilde \ell)^{1/2}$ (and not on $\ell$ and $\tilde\ell$ separately), and for $b=1$ they reduce to the expressions obtained on the sphere in \cite{Pestun:2007rz}. Moreover, as shown in \cite{Hama:2012bg}, they are symmetric in 
the exchange $b\leftrightarrow b^{-1}$. As a consequence of this symmetry, the partition function $\cZ_b$ does not depend on $b$ at first order, namely
\begin{align}
	\label{derZb}
		\partial_b\, \mathcal Z_b\,\Big|_{b=1} \!= 0~.
\end{align}
Now, the localized Wilson loop expectation value reads:
\begin{align}
	\label{derWbmm}
		\vev{ \cW_b } = \frac{1}{\cZ_b} \int\! da~ 
		\cW_b~ \mathrm{e}^{-\tr a^2} \,\big|\cZ^\mathrm{1-loop}_b\big|^2
		~ \big|\cZ^\mathrm{inst}_b\big|^2~,
\end{align}
and we can evaluate explicitly the right hand side of \eqref{conjhW}. Due to \eqref{derZb}, we get a contribution only when the $b$ derivative is applied to $\cW_b$ itself. Thus, we obtain
\begin{align}
	\label{derbBhW}
	h_W = \frac{1}{12\pi^2}	\partial_b \log \big\langle\, \mathcal W_b\,\big\rangle \Big|_{b=1}
		= \frac{1}{12\pi^2} \frac{\big\langle\partial_b \cW_b\big|_{b=1}\big\rangle\phantom{\Big|}}{\big\langle \cW\big\rangle}~.   
\end{align}
Here $\cW$ stands for $\cW_b\big|_{b=1}$.
Due to the ellipsoid properties, these vacuum expectation values are now evaluated in the $\cN=2$ matrix model on the sphere, displayed in \eqref{ZN2}. 

\paragraph{The $\cN=4$ case:} In the $\cN=4$ SYM theory,  
the matrix model is purely Gaussian as both the one-loop determinant and the instanton factor reduce to $1$. Then, observing the identity
\begin{equation}
\partial_b \cW_b\big|_{b=1} = g\,\frac{\pa \cW}{\pa g}~,
\end{equation}
and since the $g$ dependence lies purely in $\cW$, the $g$-derivative commutes with the expectation value and thus we re-obtain the result of \cite{Correa:2012at}:
\begin{equation}
	\label{N4B}
		h_W\Big|_{\cN=4}= \frac{1}{12\pi^2}\, g \partial_g \log \vev{\cW}~,
\end{equation}
which corresponds to \eqref{exactBN4}, provided that $B=3h_W$. \\
Therefore it is clear that the result \eqref{conjhW} is a generalization of the $\cN=4$ result.

\paragraph{Perturbative expansions in $\cN=2$} This big simplification no longer occurs in the $\cN=2$ case, due to the non-trivial 1-loop determinant and instanton factors. Nevertheless, the quantity in \eqref{derbBhW} can be computed in the interacting $\cN=2$ matrix model on $S^4$. In particular, employing the techniques of \cite{Billo:2018oog,Billo:2019fbi} reviewed in appendix \ref{App:mm}, we describe its perturbative expansion in $g$. 
We consider the perturbative limit in which the coupling $g$ is small and the instanton contributions
become trivial setting $\cZ_\mathrm{inst}=1$. 
The one-loop contribution can be expanded
as an interacting term:
\begin{align}\label{intAction}
		|\mathcal Z_{\mathrm{1-loop}}|^2 =~\mathrm{e}^{-S_\mathrm{int}}~,\hspace{0.7cm}	
		S_{\mathrm{int}}= \sum_{n=2}(-1)^{n}\parenth{\frac{g^2}{8\pi^2}}^{n}\,\frac{\zeta(2n-1)}{n}\Tr^\prime_\cR a^{2n}~,
\end{align}
where the combination $\Tr^\prime_\cR$ is defined in \eqref{deftrp}.
The correlator of a generic observable in the $\cN=2$ matrix model 
can be expressed in terms of vacuum expectation values computed in the Gaussian model, simply in presence of the interacting action \eqref{intAction}. We can rewrite \eqref{derbBhW} as
\begin{align}\label{hWvev0}
h_W=\frac{1}{12\pi^2}\dfrac{\vev{\partial_b\cW_b\big|_{b=1}\,\rme^{-S_\mathrm{int}}}_0}{\vev{ \cW\,\rme^{-S_\mathrm{int}}}_0}~,
\end{align}
where the subscript $0$ stands for the evaluation in the Gaussian matrix model. \\
Expanding $\cW$ and $\partial_b\cW_b\big|_{b=1}$ as well as $S_\mathrm{int}$ in series of small $g$, we get a perturbative expansion of $h_W$ in terms of expectation values of multi-trace combinations of the matrix $a$ in the Gaussian integral. Such combinations can be computed recursively at very high orders in perturbation theory and for a large class of $\cN=2$ theories defined by the representation $\cR$ of the matter hypermultiplets.
The final result can be nicely organized in terms of the Riemann zeta-values appearing in \eqref{intAction}. Such organization could be useful for studying related quantities and different theories. 

We report the first perturbative orders at low transcendentalities for the perturbative expansion of $h_W$:
\begin{align}\label{perturbativehW}
    h_W = \frac{1}{12\pi^2}\,\frac{g^2(N^2-1)}{4N} \,\parenth{1 + \zeta(3) \left(\frac{g^2}{8\pi^2}\right)^2 \cC_{4}
		- \zeta(5) \left(\frac{g^2}{8\pi^2}\right)^3 \cC_{6} + \cO(g^8)}~,
\end{align}
where $\cC_{2n}$ is the totally symmetric contraction of the $SU(N)$ tensor
$\cC_{a_1\ldots a_{2n}} = \trp T_{a_1}\ldots T_{a_{2n}}$.
In general $\cC_{2n}$ is a rational function in the number of colors $N$ (we refer to Section 3 of \cite{Billo:2019fbi} for more details). For example, in superconformal QCD theory (where $\cR$ is defined by $N_f=2N$ fundamental hypermultiplets) one finds
\begin{align}
	\label{C4C6}
		\cC_4=-3(N^2+1)~,~~~~~~
		\cC_6=-\frac{15(N^2+1)(2N^2-1)}{2 N}~.
\end{align}
Using the relations \eqref{CDtohW} and \eqref{BtohW2}, one can derive the analogous perturbative expansions of the coefficient $C_D$ in two-point function of the displacement operator and the Bremsstrahlung function $B$. These results have also been explicitly checked against some honest Feynman diagrams computations up to four-loops order in \cite{Billo:2019fbi}.
Using the sphere matrix model achievements for the Wilson loop vev (see also \cite{Andree:2010na,Fiol:2020bhf}), the computation of $h_W$ could be pushed even further in perturbation theory, and explored in the strongly coupled regime \cite{Passerini:2011fe,Fiol:2015mrp}. All these results represent a further proof of the validity of the squashed sphere formula \eqref{conjhW}.

\section{Conclusions and perspectives}\label{sec:5}
In this paper we reviewed the emitted radiation of a relativistic particle in an accelerated motion in some special classes of theories. While this problem is easily solvable for abelian gauge theories, it definitely becomes less trivial for non-abelian theories, where the addition of superconformal symmetry is needed to produce exact results in the coupling constant. The final outcome for the Bremsstrahlung function is given by equation \eqref{BdlogW}, which follows from an analogous formula \eqref{conjhW} for the stress tensor coefficient. These expressions relate the emitted radiation to a variation of the geometry of the system, following the geometric interpretation for the insertion of a stress tensor operator.

This derivation is valid for any four-dimensional superconformal theory with at least $\cN=2$ supersymmetries, and only uses general properties of the geometric background and of defect CFTs. Thus the relation \eqref{conjhW} is valid for any superconformal line defect. Furthermore, it provides a general recipe to extract exact results for the stress tensor correlators by perturbing the background geometry. Such idea could be applied to a wider class of defects \cite{Bianchi:2019sxz,Bianchi:2019umv}. We stress that this is a peculiarity of defect CFTs, where the one-point function is non-vanishing and the first-order derivative returns a non-trivial result. 

On the other hand, the extended supersymmetry allows to exploit the power of supersymmetric localization. Hence, formula \eqref{conjhW} together with the equalities \eqref{CDtohW} and \eqref{BtohW2} implies that all these apparently distinct observables are captured by a non-local operator localized on a deformed geometry. With the help of the localized matrix model, it is possible to explicitly compute perturbative expansions with a well defined organization in terms of transcendentality, which is a prediction for Feynman diagrams computations. Since in special $\cN=2$ theories it is also possible to go beyond perturbation theory \cite{Beccaria:2020hgy,Galvagno:2020cgq,Galvagno:2021bbj,Beccaria:2021hvt,Beccaria:2021ksw,Beccaria:2021vuc,Billo:2021rdb}, it would be interesting to explore the $\cN=2$ emitted radiation in these contexts to explore the holographic perspectives. 

Finally, additional directions would be to generalize the relation between $C_D$ and $h_W$ in specific setups for non supersymmetric lines \cite{Beccaria:2017rbe,Beccaria:2018ocq,Cuomo:2021rkm,Beccaria:2021rmj}, or explore the integrability approach for the generalized Bremsstrahlung \cite{Giombi:2018hsx,Giombi:2018qox,Giombi:2021zfb}.

\vskip 1.5cm
\noindent {\large {\bf Acknowledgments}}
\vskip 0.2cm
The author wants to thank L. Bianchi, M. Bill\`o, G. Cuomo, B. Fiol and A. Lerda
for carefully reading the draft and for many useful comments. \\
\noindent
This work is supported by a grant from the Swiss National Science Foundation, as well as via the NCCR SwissMAP.
\vskip 1cm
\begin{appendix}
\section{Sphere matrix model}\label{App:mm}
The localization procedure performed by Pestun \cite{Pestun:2007rz} allows to reduce the path integral of any Lagrangian supersymmetric theory with at least $\cN=2$ supersymmetry to a matrix model on a four sphere $S^4$. We review the construction of this matrix model, concentrating on a mostly perturbative approach.

We consider $\cN=2$ theories with gauge group SU($N$) and matter hypermultiplets transforming in a generic representation $\cR$ which preserves conformal symmetry. The only non-vanishing contributions to the path integral arise from the localization locus, which is defined by the following saddle points:
\begin{align}
\label{saddlepoints}
	A_\mu=0~, ~~~ \phi=\bar \phi=\,\frac{a}{\sqrt{2}}~, 
\end{align}
where $A_\mu$ is the gauge field, $\phi$ and $\bar\phi$ are the scalar fields of the vector multiplet and $a$ is a $N \times N$ Hermitean matrix which can be decomposed over a basis of generators $t_a$ of $\mathfrak{su}(N)$ Lie algebra:
\begin{equation}
	\label{aintermsoft}
	a = a^b \,t_b~,~~~b = 1,\ldots, N^2-1~.
\end{equation}
The partition function on a 4-sphere $S^4$ with unit radius can be expressed as follows:
\begin{equation}\label{ZtotS4}
	\cZ_{S^4}=\int da \,
	\big| Z_{\mathrm{tree}}\,Z_{\mathrm{1-loop}}\,
	Z_{\mathrm{inst}}\big|^2~,	
\end{equation}
where the integration measure stands for the integral over the eigenvalues of the matrix $a$.
$Z_{\mathrm{inst}}$ is the Nekrasov's instanton partition function \cite{Nekrasov:2002qd,Nekrasov:2003rj} that can be put to 1 when considering perturbation theory.\\ The tree-level term is given by a Gaussian term:
\begin{equation}
\big|Z_{\mathrm{tree}}\big|^2 = \mathrm{e}^{-\tr a^2}~.
\end{equation} 
The 1-loop determinant contains interaction terms, and can be written as:
\begin{equation}
	\label{Z1ltoexp}
	\big|Z_{\mathrm{1-loop}}\big|^2 \equiv \mathrm{e}^{-S_{\mathrm{int}}(a)}~.
\end{equation}
At this point we distinguish between the maximal supersymmetric case $\cN=4$ and generic $\cN=2$ theories.

In $\cN=4$ SYM theory all the interacting terms exactly cancel, such that $S_{\mathrm{int}}(a)=0$, and the corresponding matrix model is purely Gaussian:
\begin{equation}\label{ZN4}
    \cZ_{\cN=4} = \int da ~\mathrm{e}^{-\tr a^2}~.
\end{equation}
This fact allows to obtain several exact results in the coupling $g$.

For $\cN=2$ SYM theories interaction terms are contained in the one-loop determinant, and they explicitly depend on the matter
representation $\cR$. The perturbative matrix model can be written as follows:
\begin{equation}
    \begin{aligned}
	\label{ZN2}
		\cZ_{\cN=2} &= 	\int da~\mathrm{e}^{-\tr\, a^2 - S_{\mathrm{int}}(a)}~, \\
		S_{\mathrm{int}}(a)
		&=	-\left(\frac{g^2}{8\pi^2}\right)^2\frac{\zeta(3)}{2}\,\Tr_{\cR}^\prime a^{4}
		+\left(\frac{g^2}{8\pi^2}\right)^3\frac{\zeta(5)}{3}\,\Tr_{\cR}^\prime a^{6}
		+\ldots
\end{aligned}
\end{equation}
where $\zeta(n)$ are the Riemann $\zeta$-values and $\Tr_{\cR}^\prime$ stands for the notation:
\begin{align}
	\label{deftrp}
	\Tr_{\cR}^\prime \bullet \,= \,\Tr_{\cR} \bullet - \Tr_{\text{adj}} \bullet~.
\end{align}
From this shape we immediately see that $S_{\mathrm{int}}(a)=0$ for $\cN=4$ SYM, which can be seen as a $\cN=2$ theory with a single hypermultiplet in the adjoint representation. In $\cN=2$ theories, the matrix model \eqref{ZN2} is valid for any conformal matter content, see \cite{Billo:2019fbi} for explicit examples, and it allows for explicit perturbative computations for several supersymmetric-invariant observables.

\section{Useful formulas for the ellipsoid computation}\label{app}
We report here some useful formulas explicitly obtained in \cite{Bianchi:2019dlw} that are necessary to perform the ellipsoid integrals discussed in section \ref{sec.4.2}.

\subsection{Background values for the supergravity multiplet}\label{app:1}
The Killing spinor equations provide specific geometric constraints that allow to fix the 
profile of the background fields, although not uniquely. This arbitrariness is fully discussed in \cite{Hama:2012bg} and does not affect the final result of the calculation. Due to \eqref{tmujmu0} we can avoid to write the results for $V_\m^0$ and $(V_\mu)^\cI_\cJ$. 
All the results are written in terms of three functions:
\begin{align}\label{f1f2f3}
    f_1 \!=\!\sqrt{\ell^2 \sin^2 \theta+\widetilde\ell^{\,2}\cos^2 \theta}~,~~~
		f_2 \!=\! \sqrt{r^2\sin^2 \rho+ \frac{1}{f^2}\ell^2\widetilde\ell^{\,2} \cos^2 \rho}~,~~~
		f_3 \!=\! \frac{\widetilde\ell^{\,2}-\ell^2}{f}\cos \rho \sin \theta \cos \theta~.
\end{align}
The solution for the metric is given in terms of the vierbeins $E^m$, while the (anti-)self-dual
tensors $\mathsf{T}_{\mu\nu}$ and $\bar{\mathsf{T}}_{\mu\nu}$ are related to the matrices 
$\mathsf{T}_{\alpha}^{\beta} $ and $\bar{\mathsf{T}}^{\dot \alpha}_{\dot \beta} $ as:
\begin{align}
	\label{Tab}
		\mathsf{T}_{\alpha}^{\beta} = -\ii\parenth{\sigma^{\mu\nu}}_{\alpha}^{\beta}\, \mathsf{T}_{\mu\nu}~,~~~~
		\bar{\mathsf{T}}^{\dot \alpha}_{\dot \beta} =-\ii \parenth{\bar{\sigma}^{\mu\nu}}^{\dot \alpha}_{\dot \beta} \,\bar{\mathsf{T}}_{\mu\nu}~.
\end{align}
The explicit solutions read:
\begin{subequations}
\label{Backgroundsol}
\begin{align}	
 M & =\frac{1}{f_1^2}+\frac{f_3^2+r^2}{f_1^2f_2^2}-\frac{4}{f_1f_2}~,\label{Msol}\\[2mm]
\mathsf{T}_{\a}^{\b} & = \frac{1}{4}\parenth{ \frac{1}{f_1}-\frac{1}{f_2}} (\t_{\theta}^1)_{\a}^{\b} 
		+\frac{f_3}{4 f_1f_2}\, (\t_{\theta}^2)_{\a}^{\b}~,\label{Tsol}\\[2mm]
\bar{\mathsf{T}}^{\dot \a}_{\dot \b} & = \frac{1}{4}\parenth{ \frac{1}{f_1}-\frac{1}{f_2}}
		(\t_{\theta}^1)^{\dot \a}_{\dot \b} -\frac{f_3}{4 f_1 f_2}\, (\t_{\theta}^2)^{\dot \a}_{\dot \b}~,\label{barTsol}\\[2mm]
		E^1 &= \ell\, \sin \rho \cos \theta\, d\varphi~,\quad\quad\quad
		E^2 = \widetilde{\ell}\,\sin \rho \sin \theta\, d\chi~,\notag \\ 
		E^3 &= f_1 \,\sin \rho\, d\theta+f_3\, d\rho~,\quad\quad
		E^4 = f_2\, d\rho~,\label{vierbein}
		\end{align}
\end{subequations}
where the matrices $\tau^i_\theta$ are
\begin{align}\label{tthetamat}
		\tau^i_\theta = \tau^i \,\begin{pmatrix} \mathrm{e}^{\ii \theta} &0  \\ 0& \mathrm{e}^{-\ii \theta} 
		\end{pmatrix} ~,
\end{align}
and $\tau^i$ are the Pauli matrices. 

\subsection{One-point functions of the stress tensor multiplet}\label{app:2}
We report the explicit results of the one-point functions of the components of the stress tensor multiplet which are necessary to perform the ellipsoid integration.
We start from the stress tensor one-point function $\vev{T^{\mu\nu}}_W$. To write the kinematic structure of a correlator with spin, one can contract 
all indices with a complex vector $z_\mu$, such that $ z^\mu g^0_{\mu\nu} z^{\nu}=0$. 
Then the one-point function of this tensor is a polynomial in $z$:
\begin{align}\label{Tmunu1pt}
z^{\mu} z^{\nu} \big\langle T_{\mu\nu}\big\rangle_W
\!=h_W\,\frac{z_\chi^2\,\sin^2\theta \sin ^2\rho 
		 \big( \cos 2 \theta\!-\!2\cos ^2\theta \cos 2 \rho \!-\!3\big)
		 \!-\!4\big(z_\rho\,\sin\theta \!+\!z_\theta\,\cos \theta\sin \rho \cos \rho\big)^2}{\mathsf{r}^4 \big(\cos ^2\rho+\sin ^2\theta
		   \sin ^2\rho\big)^3}~.
\end{align}
To open indices again, one can apply
the Todorov operator to this polynomial
\begin{align}
\label{todorov}
\cD_{\mu}=\parenth{1+z^\lambda \frac{\partial}{\partial z^\lambda} }\frac{\partial}{\partial z^{\mu}}-\frac{1}{2} z_{\mu} \frac{\partial^2}{\partial z^\rho  \partial z_\rho}~.
\end{align}
Similarly, for the (anti-)self dual operators the one-point functions read:
\begin{align}\label{HbarHsol}
\vev{H_{\a}^{\b}}_W &=\frac{3\ii h_W}{4}\,\frac{\cos\theta \cos\rho \,(\tau^1)_{\alpha}^{\beta}
-\sin\theta \,(\tau^2)_{\alpha}^{\beta}}{\mathsf{r}^3
 \parenth{\cos ^2\rho+\sin ^2\theta \sin ^2\rho}^2}~,\notag \\[2mm]
\vev{\bar{H}^{\dot \alpha}_{\dot\beta}}_W &
=-\frac{3\ii h_W}{4}\,\frac{\cos\theta \cos\rho \,(\tau^1)^{\dot\a}_{\dot\b}
+\sin\theta \,(\tau^2)^{\dot\a}_{\dot\b}}{\mathsf{r}^3
 \parenth{\cos ^2\rho+\sin ^2\theta \sin ^2\rho}^2}~.
\end{align}
The last one-point function that we need is the scalar superprimary operator
$O_2$:
\begin{align}\label{O2sol}
		\vev{O_2}_W=\frac{3h_W}{8}\,
		\frac{1}{\mathsf{r}^2 \big(\cos^2 \rho+ \sin^2 \theta \sin^2 \rho\big)}~.
\end{align}
The coefficients for $\vev{H_{\mu\nu}}_W$, $\vev{\bar H_{\mu\nu}}_W$ and $\vev{O_2}_W$ have been fixed in terms of $h_W$ using superconformal Ward identities.

\end{appendix}

\bibliography{biblio_Tesi}

\providecommand{\href}[2]{#2}\begingroup\raggedright\begin{thebibliography}{10}

\bibitem{Dirac:1938nz}
P.~A.~M. Dirac, \emph{{Classical theory of radiating electrons}},
\href{http://dx.doi.org/10.1098/rspa.1938.0124}{Proc. Roy. Soc. Lond. {\bf
  A167} (1938)  148--169}.

\bibitem{Jackson}
J.~Jackson, {\em {Classical Electrodynamics}}.
\newblock John Wiley \& Sons, Inc., 1975.

\bibitem{Landau:1975pou}
L.~D. Landau and E.~M. Lifschits, {\em {The Classical Theory of Fields}},
  vol.~Volume 2 of {\em Course of Theoretical Physics}.
\newblock Pergamon Press, Oxford, 1975.

\bibitem{Polyakov:1980ca}
A.~M. Polyakov, \emph{{Gauge Fields as Rings of Glue}},
\href{http://dx.doi.org/10.1016/0550-3213(80)90507-6}{Nucl. Phys. {\bf B164}
  (1980)  171--188}.

\bibitem{Correa:2012at}
D.~Correa, J.~Henn, J.~Maldacena, and A.~Sever, \emph{{An exact formula for the
  radiation of a moving quark in N=4 super Yang Mills}},
  \href{http://dx.doi.org/10.1007/JHEP06(2012)048}{JHEP {\bf 06} (2012)  048},
\href{http://arxiv.org/abs/1202.4455}{{\tt arXiv:1202.4455 [hep-th]}}.

\bibitem{Mikhailov:2003er}
A.~Mikhailov, \emph{{Nonlinear waves in AdS / CFT correspondence}},
\href{http://arxiv.org/abs/hep-th/0305196}{{\tt arXiv:hep-th/0305196
  [hep-th]}}.

\bibitem{Athanasiou:2010pv}
C.~Athanasiou, P.~M. Chesler, H.~Liu, D.~Nickel, and K.~Rajagopal,
  \emph{{Synchrotron radiation in strongly coupled conformal field theories}},
  \href{http://dx.doi.org/10.1103/PhysRevD.81.126001,
  10.1103/PhysRevD.84.069901}{Phys. Rev. {\bf D81} (2010)  126001},
  \href{http://arxiv.org/abs/1001.3880}{{\tt arXiv:1001.3880 [hep-th]}}.
[Erratum: Phys. Rev.D84,069901(2011)].

\bibitem{Hatta:2011gh}
Y.~Hatta, E.~Iancu, A.~H. Mueller, and D.~N. Triantafyllopoulos,
  \emph{{Radiation by a heavy quark in N=4 SYM at strong coupling}},
  \href{http://dx.doi.org/10.1016/j.nuclphysb.2011.04.011}{Nucl. Phys. B {\bf
  850} (2011)  31--52}, \href{http://arxiv.org/abs/1102.0232}{{\tt
  arXiv:1102.0232 [hep-th]}}.

\bibitem{Correa:2012hh}
D.~Correa, J.~Maldacena, and A.~Sever, \emph{{The quark anti-quark potential
  and the cusp anomalous dimension from a TBA equation}},
  \href{http://dx.doi.org/10.1007/JHEP08(2012)134}{JHEP {\bf 08} (2012)  134},
\href{http://arxiv.org/abs/1203.1913}{{\tt arXiv:1203.1913 [hep-th]}}.

\bibitem{Drukker:2012de}
N.~Drukker, \emph{{Integrable Wilson loops}},
  \href{http://dx.doi.org/10.1007/JHEP10(2013)135}{JHEP {\bf 10} (2013)  135},
\href{http://arxiv.org/abs/1203.1617}{{\tt arXiv:1203.1617 [hep-th]}}.

\bibitem{Gromov:2012eu}
N.~Gromov and A.~Sever, \emph{{Analytic Solution of Bremsstrahlung TBA}},
  \href{http://dx.doi.org/10.1007/JHEP11(2012)075}{JHEP {\bf 11} (2012)  075},
  \href{http://arxiv.org/abs/1207.5489}{{\tt arXiv:1207.5489 [hep-th]}}.

\bibitem{Gromov:2013qga}
N.~Gromov, F.~Levkovich-Maslyuk, and G.~Sizov, \emph{{Analytic Solution of
  Bremsstrahlung TBA II: Turning on the Sphere Angle}},
  \href{http://dx.doi.org/10.1007/JHEP10(2013)036}{JHEP {\bf 10} (2013)  036},
\href{http://arxiv.org/abs/1305.1944}{{\tt arXiv:1305.1944 [hep-th]}}.

\bibitem{Billo:2016cpy}
M.~Billo, V.~Goncalves, E.~Lauria, and M.~Meineri, \emph{{Defects in conformal
  field theory}}, \href{http://dx.doi.org/10.1007/JHEP04(2016)091}{JHEP {\bf
  04} (2016)  091},
\href{http://arxiv.org/abs/1601.02883}{{\tt arXiv:1601.02883 [hep-th]}}.

\bibitem{Fiol:2012sg}
B.~Fiol, B.~Garolera, and A.~Lewkowycz, \emph{{Exact results for static and
  radiative fields of a quark in N=4 super Yang-Mills}},
  \href{http://dx.doi.org/10.1007/JHEP05(2012)093}{JHEP {\bf 05} (2012)  093},
\href{http://arxiv.org/abs/1202.5292}{{\tt arXiv:1202.5292 [hep-th]}}.

\bibitem{Lewkowycz:2013laa}
A.~Lewkowycz and J.~Maldacena, \emph{{Exact results for the entanglement
  entropy and the energy radiated by a quark}},
  \href{http://dx.doi.org/10.1007/JHEP05(2014)025}{JHEP {\bf 05} (2014)  025},
\href{http://arxiv.org/abs/1312.5682}{{\tt arXiv:1312.5682 [hep-th]}}.

\bibitem{Kapustin:2005py}
A.~Kapustin, \emph{{Wilson-'t Hooft operators in four-dimensional gauge
  theories and S-duality}},
  \href{http://dx.doi.org/10.1103/PhysRevD.74.025005}{Phys. Rev. {\bf D74}
  (2006)  025005},
\href{http://arxiv.org/abs/hep-th/0501015}{{\tt arXiv:hep-th/0501015
  [hep-th]}}.

\bibitem{Fiol:2011zg}
B.~Fiol and B.~Garolera, \emph{{Energy Loss of an Infinitely Massive
  Half-Bogomol'nyi-Prasad-Sommerfeld Particle by Radiation to All Orders in
  $1/N$}}, \href{http://dx.doi.org/10.1103/PhysRevLett.107.151601}{Phys. Rev.
  Lett. {\bf 107} (2011)  151601},
\href{http://arxiv.org/abs/1106.5418}{{\tt arXiv:1106.5418 [hep-th]}}.

\bibitem{Fiol:2013hna}
B.~Fiol and G.~Torrents, \emph{{Exact results for Wilson loops in arbitrary
  representations}}, \href{http://dx.doi.org/10.1007/JHEP01(2014)020}{JHEP {\bf
  01} (2014)  020}, \href{http://arxiv.org/abs/1311.2058}{{\tt arXiv:1311.2058
  [hep-th]}}.

\bibitem{Fiol:2014vqa}
B.~Fiol, A.~Güijosa, and J.~F. Pedraza, \emph{{Branes from Light: Embeddings
  and Energetics for Symmetric $k$-Quarks in $\mathcal{N}=4$ SYM}},
  \href{http://dx.doi.org/10.1007/JHEP01(2015)149}{JHEP {\bf 01} (2015)  149},
\href{http://arxiv.org/abs/1410.0692}{{\tt arXiv:1410.0692 [hep-th]}}.

\bibitem{Forini:2012bb}
V.~Forini, V.~G.~M. Puletti, and O.~Ohlsson~Sax, \emph{{The generalized cusp in
  $AdS_4 \times CP^3$ and more one-loop results from semiclassical strings}},
  \href{http://dx.doi.org/10.1088/1751-8113/46/11/115402}{J. Phys. {\bf A46}
  (2013)  115402},
\href{http://arxiv.org/abs/1204.3302}{{\tt arXiv:1204.3302 [hep-th]}}.

\bibitem{Correa:2014aga}
D.~H. Correa, J.~Aguilera-Damia, and G.~A. Silva, \emph{{Strings in $AdS_4
  \times \mathbb{CP}^{3}$ Wilson loops in $\mathcal N=$6 super
  Chern-Simons-matter and bremsstrahlung functions}},
  \href{http://dx.doi.org/10.1007/JHEP06(2014)139}{JHEP {\bf 06} (2014)  139},
\href{http://arxiv.org/abs/1405.1396}{{\tt arXiv:1405.1396 [hep-th]}}.

\bibitem{Bianchi:2014ada}
L.~Bianchi, M.~S. Bianchi, A.~Bres, V.~Forini, and E.~Vescovi, \emph{{Two-loop
  cusp anomaly in ABJM at strong coupling}},
  \href{http://dx.doi.org/10.1007/JHEP10(2014)013}{JHEP {\bf 10} (2014)  013},
\href{http://arxiv.org/abs/1407.4788}{{\tt arXiv:1407.4788 [hep-th]}}.

\bibitem{Bianchi:2014laa}
M.~S. Bianchi, L.~Griguolo, M.~Leoni, S.~Penati, and D.~Seminara, \emph{{BPS
  Wilson loops and Bremsstrahlung function in ABJ(M): a two loop analysis}},
  \href{http://dx.doi.org/10.1007/JHEP06(2014)123}{JHEP {\bf 06} (2014)  123},
\href{http://arxiv.org/abs/1402.4128}{{\tt arXiv:1402.4128 [hep-th]}}.

\bibitem{Bianchi:2017ozk}
L.~Bianchi, L.~Griguolo, M.~Preti, and D.~Seminara, \emph{{Wilson lines as
  superconformal defects in ABJM theory: a formula for the emitted radiation}},
  \href{http://dx.doi.org/10.1007/JHEP10(2017)050}{JHEP {\bf 10} (2017)  050},
\href{http://arxiv.org/abs/1706.06590}{{\tt arXiv:1706.06590 [hep-th]}}.

\bibitem{Bianchi:2017svd}
M.~S. Bianchi, L.~Griguolo, A.~Mauri, S.~Penati, M.~Preti, and D.~Seminara,
  \emph{{Towards the exact Bremsstrahlung function of ABJM theory}},
  \href{http://dx.doi.org/10.1007/JHEP08(2017)022}{JHEP {\bf 08} (2017)  022},
\href{http://arxiv.org/abs/1705.10780}{{\tt arXiv:1705.10780 [hep-th]}}.

\bibitem{Bianchi:2018bke}
M.~S. Bianchi, L.~Griguolo, A.~Mauri, S.~Penati, and D.~Seminara, \emph{{A
  matrix model for the latitude Wilson loop in ABJM theory}},
\href{http://arxiv.org/abs/1802.07742}{{\tt arXiv:1802.07742 [hep-th]}}.

\bibitem{Bianchi:2018scb}
L.~Bianchi, M.~Preti, and E.~Vescovi, \emph{{Exact Bremsstrahlung functions in
  ABJM theory}}, \href{http://dx.doi.org/10.1007/JHEP07(2018)060}{JHEP {\bf 07}
  (2018)  060},
\href{http://arxiv.org/abs/1802.07726}{{\tt arXiv:1802.07726 [hep-th]}}.

\bibitem{Drukker:2019bev}
N.~Drukker {\em et al.}, \emph{{Roadmap on Wilson loops in 3d
  Chern\textendash{}Simons-matter theories}},
  \href{http://dx.doi.org/10.1088/1751-8121/ab5d50}{J. Phys. A {\bf 53} (2020)
  no.~17, 173001}, \href{http://arxiv.org/abs/1910.00588}{{\tt arXiv:1910.00588
  [hep-th]}}.

\bibitem{Fiol:2015spa}
B.~Fiol, E.~Gerchkovitz, and Z.~Komargodski, \emph{{Exact Bremsstrahlung
  Function in $N=2$ Superconformal Field Theories}},
  \href{http://dx.doi.org/10.1103/PhysRevLett.116.081601}{Phys. Rev. Lett. {\bf
  116} (2016) no.~8, 081601},
\href{http://arxiv.org/abs/1510.01332}{{\tt arXiv:1510.01332 [hep-th]}}.

\bibitem{Mitev:2014yba}
V.~Mitev and E.~Pomoni, \emph{{Exact effective couplings of four dimensional
  gauge theories with $\mathcal N=$ 2 supersymmetry}},
  \href{http://dx.doi.org/10.1103/PhysRevD.92.125034}{Phys. Rev. {\bf D92}
  (2015) no.~12, 125034},
\href{http://arxiv.org/abs/1406.3629}{{\tt arXiv:1406.3629 [hep-th]}}.

\bibitem{Mitev:2015oty}
V.~Mitev and E.~Pomoni, \emph{{Exact Bremsstrahlung and Effective Couplings}},
  \href{http://dx.doi.org/10.1007/JHEP06(2016)078}{JHEP {\bf 06} (2016)  078},
\href{http://arxiv.org/abs/1511.02217}{{\tt arXiv:1511.02217 [hep-th]}}.

\bibitem{Gomez:2018usu}
C.~Gomez, A.~Mauri, and S.~Penati, \emph{{The Bremsstrahlung function of $
  \mathcal{N} $ = 2 SCQCD}},
  \href{http://dx.doi.org/10.1007/JHEP03(2019)122}{JHEP {\bf 03} (2019)  122},
\href{http://arxiv.org/abs/1811.08437}{{\tt arXiv:1811.08437 [hep-th]}}.

\bibitem{Bianchi:2018zpb}
L.~Bianchi, M.~Lemos, and M.~Meineri, \emph{{Line Defects and Radiation in
  $\mathcal{N}=2$ Conformal Theories}},
  \href{http://dx.doi.org/10.1103/PhysRevLett.121.141601}{Phys. Rev. Lett. {\bf
  121} (2018) no.~14, 141601},
\href{http://arxiv.org/abs/1805.04111}{{\tt arXiv:1805.04111 [hep-th]}}.

\bibitem{Bianchi:2019dlw}
L.~Bianchi, M.~Billo, F.~Galvagno, and A.~Lerda, \emph{{Emitted Radiation and
  Geometry}}, \href{http://dx.doi.org/10.1007/JHEP01(2020)075}{JHEP {\bf 01}
  (2020)  075},
\href{http://arxiv.org/abs/1910.06332}{{\tt arXiv:1910.06332 [hep-th]}}.

\bibitem{Grozin:2015kna}
A.~Grozin, J.~M. Henn, G.~P. Korchemsky, and P.~Marquard, \emph{{The three-loop
  cusp anomalous dimension in QCD and its supersymmetric extensions}},
  \href{http://dx.doi.org/10.1007/JHEP01(2016)140}{JHEP {\bf 01} (2016)  140},
\href{http://arxiv.org/abs/1510.07803}{{\tt arXiv:1510.07803 [hep-ph]}}.

\bibitem{Schild:1960}
A.~Schild, \emph{{On the Radiation Emitted by an Accelerated Point Charge}},
  Riv. Nuovo Cim. {\bf 1} (1960)  127--131.

\bibitem{Teitelboim:1979px}
C.~Teitelboim, D.~Villarroel, and C.~van Weert, \emph{{Classical
  Electrodynamics of Retarded Fields and Point Particles}},
\href{http://dx.doi.org/10.1007/BF02895735}{Riv. Nuovo Cim. {\bf 3N9} (1980)
  1--64}.

\bibitem{Rohrlich}
F.~Rohrlich, {\em {Classical Charged Particles}}.
\newblock World Scientific Publishing Company, Singapore, 2007.

\bibitem{Fiol:2019woe}
B.~Fiol and J.~Mart\'\i{}nez-Montoya, \emph{{On scalar radiation}},
  \href{http://dx.doi.org/10.1007/JHEP03(2020)087}{JHEP {\bf 03} (2020)  087},
  \href{http://arxiv.org/abs/1907.08161}{{\tt arXiv:1907.08161 [hep-th]}}.

\bibitem{Fulton}
T.~Fulton and F.~Rohrlich,
  \href{http://dx.doi.org/10.1016/0003-4916(60)90105-6}{\emph{{Classical
  radiation from a uniformly accelerated charge}},Annals of Physics {\bf 9}
  (Apr., 1960)  499--517}.

\bibitem{Boulware:1979qj}
D.~G. Boulware, \emph{{Radiation From a Uniformly Accelerated Charge}},
\href{http://dx.doi.org/10.1016/0003-4916(80)90360-7}{Annals Phys. {\bf 124}
  (1980)  169}.

\bibitem{Erickson:2000af}
J.~K. Erickson, G.~W. Semenoff, and K.~Zarembo, \emph{{Wilson loops in N=4
  supersymmetric Yang-Mills theory}},
  \href{http://dx.doi.org/10.1016/S0550-3213(00)00300-X}{Nucl. Phys. {\bf B582}
  (2000)  155--175},
\href{http://arxiv.org/abs/hep-th/0003055}{{\tt arXiv:hep-th/0003055
  [hep-th]}}.

\bibitem{Drukker:2000rr}
N.~Drukker and D.~J. Gross, \emph{{An Exact prediction of N=4 SUSYM theory for
  string theory}}, \href{http://dx.doi.org/10.1063/1.1372177}{J. Math. Phys.
  {\bf 42} (2001)  2896--2914},
\href{http://arxiv.org/abs/hep-th/0010274}{{\tt arXiv:hep-th/0010274
  [hep-th]}}.

\bibitem{Pestun:2007rz}
V.~Pestun, \emph{{Localization of gauge theory on a four-sphere and
  supersymmetric Wilson loops}},
  \href{http://dx.doi.org/10.1007/s00220-012-1485-0}{Commun. Math. Phys. {\bf
  313} (2012)  71--129},
\href{http://arxiv.org/abs/0712.2824}{{\tt arXiv:0712.2824 [hep-th]}}.

\bibitem{Pestun:2016zxk}
V.~Pestun {\em et al.}, \emph{{Localization techniques in quantum field
  theories}}, \href{http://dx.doi.org/10.1088/1751-8121/aa63c1}{J. Phys. {\bf
  A50} (2017) no.~44, 440301},
\href{http://arxiv.org/abs/1608.02952}{{\tt arXiv:1608.02952 [hep-th]}}.

\bibitem{Drukker:2006ga}
N.~Drukker, \emph{{1/4 BPS circular loops, unstable world-sheet instantons and
  the matrix model}},
  \href{http://dx.doi.org/10.1088/1126-6708/2006/09/004}{JHEP {\bf 09} (2006)
  004},
\href{http://arxiv.org/abs/hep-th/0605151}{{\tt arXiv:hep-th/0605151
  [hep-th]}}.

\bibitem{Drukker:2006zk}
N.~Drukker, S.~Giombi, R.~Ricci, and D.~Trancanelli, \emph{{On the D3-brane
  description of some 1/4 BPS Wilson loops}},
  \href{http://dx.doi.org/10.1088/1126-6708/2007/04/008}{JHEP {\bf 04} (2007)
  008},
\href{http://arxiv.org/abs/hep-th/0612168}{{\tt arXiv:hep-th/0612168
  [hep-th]}}.

\bibitem{Drukker:2007dw}
N.~Drukker, S.~Giombi, R.~Ricci, and D.~Trancanelli, \emph{{More supersymmetric
  Wilson loops}}, \href{http://dx.doi.org/10.1103/PhysRevD.76.107703}{Phys.
  Rev. {\bf D76} (2007)  107703},
\href{http://arxiv.org/abs/0704.2237}{{\tt arXiv:0704.2237 [hep-th]}}.

\bibitem{Drukker:2007qr}
N.~Drukker, S.~Giombi, R.~Ricci, and D.~Trancanelli, \emph{{Supersymmetric
  Wilson loops on S**3}},
  \href{http://dx.doi.org/10.1088/1126-6708/2008/05/017}{JHEP {\bf 05} (2008)
  017},
\href{http://arxiv.org/abs/0711.3226}{{\tt arXiv:0711.3226 [hep-th]}}.

\bibitem{Drukker:2007yx}
N.~Drukker, S.~Giombi, R.~Ricci, and D.~Trancanelli, \emph{{Wilson loops: From
  four-dimensional SYM to two-dimensional YM}},
  \href{http://dx.doi.org/10.1103/PhysRevD.77.047901}{Phys. Rev. {\bf D77}
  (2008)  047901},
\href{http://arxiv.org/abs/0707.2699}{{\tt arXiv:0707.2699 [hep-th]}}.

\bibitem{Pestun:2009nn}
V.~Pestun, \emph{{Localization of the four-dimensional N=4 SYM to a two-sphere
  and 1/8 BPS Wilson loops}},
  \href{http://dx.doi.org/10.1007/JHEP12(2012)067}{JHEP {\bf 12} (2012)  067},
\href{http://arxiv.org/abs/0906.0638}{{\tt arXiv:0906.0638 [hep-th]}}.

\bibitem{Giombi:2009ms}
S.~Giombi, V.~Pestun, and R.~Ricci, \emph{{Notes on supersymmetric Wilson loops
  on a two-sphere}}, \href{http://dx.doi.org/10.1007/JHEP07(2010)088}{JHEP {\bf
  07} (2010)  088},
\href{http://arxiv.org/abs/0905.0665}{{\tt arXiv:0905.0665 [hep-th]}}.

\bibitem{Giombi:2009ds}
S.~Giombi and V.~Pestun, \emph{{Correlators of local operators and 1/8 BPS
  Wilson loops on S**2 from 2d YM and matrix models}},
  \href{http://dx.doi.org/10.1007/JHEP10(2010)033}{JHEP {\bf 10} (2010)  033},
\href{http://arxiv.org/abs/0906.1572}{{\tt arXiv:0906.1572 [hep-th]}}.

\bibitem{Drukker:1999zq}
N.~Drukker, D.~J. Gross, and H.~Ooguri, \emph{{Wilson loops and minimal
  surfaces}}, \href{http://dx.doi.org/10.1103/PhysRevD.60.125006}{Phys. Rev.
  {\bf D60} (1999)  125006},
\href{http://arxiv.org/abs/hep-th/9904191}{{\tt arXiv:hep-th/9904191
  [hep-th]}}.

\bibitem{Correa:2012nk}
D.~Correa, J.~Henn, J.~Maldacena, and A.~Sever, \emph{{The cusp anomalous
  dimension at three loops and beyond}},
  \href{http://dx.doi.org/10.1007/JHEP05(2012)098}{JHEP {\bf 05} (2012)  098},
\href{http://arxiv.org/abs/1203.1019}{{\tt arXiv:1203.1019 [hep-th]}}.

\bibitem{Drukker:2011za}
N.~Drukker and V.~Forini, \emph{{Generalized quark-antiquark potential at weak
  and strong coupling}}, \href{http://dx.doi.org/10.1007/JHEP06(2011)131}{JHEP
  {\bf 06} (2011)  131},
\href{http://arxiv.org/abs/1105.5144}{{\tt arXiv:1105.5144 [hep-th]}}.

\bibitem{Fiol:2013iaa}
B.~Fiol, B.~Garolera, and G.~Torrents, \emph{{Exact momentum fluctuations of an
  accelerated quark in N=4 super Yang-Mills}},
  \href{http://dx.doi.org/10.1007/JHEP06(2013)011}{JHEP {\bf 06} (2013)  011},
  \href{http://arxiv.org/abs/1302.6991}{{\tt arXiv:1302.6991 [hep-th]}}.

\bibitem{Gomis:2008qa}
J.~Gomis, S.~Matsuura, T.~Okuda, and D.~Trancanelli, \emph{{Wilson loop
  correlators at strong coupling: From matrices to bubbling geometries}},
  \href{http://dx.doi.org/10.1088/1126-6708/2008/08/068}{JHEP {\bf 08} (2008)
  068},
\href{http://arxiv.org/abs/0807.3330}{{\tt arXiv:0807.3330 [hep-th]}}.

\bibitem{Gerchkovitz:2016gxx}
E.~Gerchkovitz, J.~Gomis, N.~Ishtiaque, A.~Karasik, Z.~Komargodski, and S.~S.
  Pufu, \emph{{Correlation Functions of Coulomb Branch Operators}},
  \href{http://dx.doi.org/10.1007/JHEP01(2017)103}{JHEP {\bf 01} (2017)  103},
\href{http://arxiv.org/abs/1602.05971}{{\tt arXiv:1602.05971 [hep-th]}}.

\bibitem{Rodriguez-Gomez:2016cem}
D.~Rodriguez-Gomez and J.~G. Russo, \emph{{Operator mixing in large $N$
  superconformal field theories on S$^{4}$ and correlators with Wilson loops}},
  \href{http://dx.doi.org/10.1007/JHEP12(2016)120}{JHEP {\bf 12} (2016)  120},
\href{http://arxiv.org/abs/1607.07878}{{\tt arXiv:1607.07878 [hep-th]}}.

\bibitem{Rodriguez-Gomez:2016ijh}
D.~Rodriguez-Gomez and J.~G. Russo, \emph{{Large N Correlation Functions in
  Superconformal Field Theories}},
  \href{http://dx.doi.org/10.1007/JHEP06(2016)109}{JHEP {\bf 06} (2016)  109},
\href{http://arxiv.org/abs/1604.07416}{{\tt arXiv:1604.07416 [hep-th]}}.

\bibitem{Billo:2017glv}
M.~Billo, F.~Fucito, A.~Lerda, J.~F. Morales, {\relax Ya}.~S. Stanev, and
  C.~Wen, \emph{{Two-point Correlators in N=2 Gauge Theories}},
  \href{http://dx.doi.org/10.1016/j.nuclphysb.2017.11.003}{Nucl. Phys. {\bf
  B926} (2018)  427--466},
\href{http://arxiv.org/abs/1705.02909}{{\tt arXiv:1705.02909 [hep-th]}}.

\bibitem{Semenoff:2001xp}
G.~W. Semenoff and K.~Zarembo, \emph{{More exact predictions of SUSYM for
  string theory}}, \href{http://dx.doi.org/10.1016/S0550-3213(01)00455-2}{Nucl.
  Phys. {\bf B616} (2001)  34--46},
\href{http://arxiv.org/abs/hep-th/0106015}{{\tt arXiv:hep-th/0106015
  [hep-th]}}.

\bibitem{Pestun:2002mr}
V.~Pestun and K.~Zarembo, \emph{{Comparing strings in AdS(5) x S**5 to planar
  diagrams: An Example}},
  \href{http://dx.doi.org/10.1103/PhysRevD.67.086007}{Phys. Rev. {\bf D67}
  (2003)  086007},
\href{http://arxiv.org/abs/hep-th/0212296}{{\tt arXiv:hep-th/0212296
  [hep-th]}}.

\bibitem{Billo:2018oog}
M.~Billo, F.~Galvagno, P.~Gregori, and A.~Lerda, \emph{{Correlators between
  Wilson loop and chiral operators in $ \mathcal{N}=2 $ conformal gauge
  theories}}, \href{http://dx.doi.org/10.1007/JHEP03(2018)193}{JHEP {\bf 03}
  (2018)  193},
\href{http://arxiv.org/abs/1802.09813}{{\tt arXiv:1802.09813 [hep-th]}}.

\bibitem{Dolan:2002zh}
F.~A. Dolan and H.~Osborn, \emph{{On short and semi-short representations for
  four-dimensional superconformal symmetry}},
  \href{http://dx.doi.org/10.1016/S0003-4916(03)00074-5}{Annals Phys. {\bf 307}
  (2003)  41--89},
\href{http://arxiv.org/abs/hep-th/0209056}{{\tt arXiv:hep-th/0209056
  [hep-th]}}.

\bibitem{Fiol:2018yuc}
B.~Fiol, J.~Mart\'\i{}nez-Montoya, and A.~Rios~Fukelman, \emph{{Wilson loops in
  terms of color invariants}},
  \href{http://dx.doi.org/10.1007/JHEP05(2019)202}{JHEP {\bf 05} (2019)  202},
  \href{http://arxiv.org/abs/1812.06890}{{\tt arXiv:1812.06890 [hep-th]}}.

\bibitem{Bianchi:2019sxz}
L.~Bianchi and M.~Lemos, \emph{{Superconformal surfaces in four dimensions}},
  \href{http://dx.doi.org/10.1007/JHEP06(2020)056}{JHEP {\bf 06} (2020)  056},
  \href{http://arxiv.org/abs/1911.05082}{{\tt arXiv:1911.05082 [hep-th]}}.

\bibitem{Hama:2012bg}
N.~Hama and K.~Hosomichi, \emph{{Seiberg-Witten Theories on Ellipsoids}},
  \href{http://dx.doi.org/10.1007/JHEP09(2012)033,
  10.1007/JHEP10(2012)051}{JHEP {\bf 09} (2012)  033},
  \href{http://arxiv.org/abs/1206.6359}{{\tt arXiv:1206.6359 [hep-th]}}.
[Addendum: JHEP10,051(2012)].

\bibitem{Festuccia:2011ws}
G.~Festuccia and N.~Seiberg, \emph{{Rigid Supersymmetric Theories in Curved
  Superspace}}, \href{http://dx.doi.org/10.1007/JHEP06(2011)114}{JHEP {\bf 06}
  (2011)  114},
\href{http://arxiv.org/abs/1105.0689}{{\tt arXiv:1105.0689 [hep-th]}}.

\bibitem{Klare:2013dka}
C.~Klare and A.~Zaffaroni, \emph{{Extended Supersymmetry on Curved Spaces}},
  \href{http://dx.doi.org/10.1007/JHEP10(2013)218}{JHEP {\bf 10} (2013)  218},
\href{http://arxiv.org/abs/1308.1102}{{\tt arXiv:1308.1102 [hep-th]}}.

\bibitem{Freedman:2012zz}
D.~Z. Freedman and A.~Van~Proeyen, {\em {Supergravity}}.
\newblock Cambridge Univ. Press, Cambridge, UK, 2012.
\newblock
\url{http://www.cambridge.org/mw/academic/subjects/physics/theoretical-physics-and-mathematical-physics/supergravity?format=AR}.
\newblock

\bibitem{Lauria:2018klo}
E.~Lauria, M.~Meineri, and E.~Trevisani, \emph{{Spinning operators and defects
  in conformal field theory}},
  \href{http://dx.doi.org/10.1007/JHEP08(2019)066}{JHEP {\bf 08} (2019)  066},
\href{http://arxiv.org/abs/1807.02522}{{\tt arXiv:1807.02522 [hep-th]}}.

\bibitem{Billo:2019fbi}
M.~Billo, F.~Galvagno, and A.~Lerda, \emph{{BPS Wilson loops in generic
  conformal $ \mathcal{N} $ = 2 SU(N) SYM theories}},
  \href{http://dx.doi.org/10.1007/JHEP08(2019)108}{JHEP {\bf 08} (2019)  108},
\href{http://arxiv.org/abs/1906.07085}{{\tt arXiv:1906.07085 [hep-th]}}.

\bibitem{Andree:2010na}
R.~Andree and D.~Young, \emph{{Wilson Loops in N=2 Superconformal Yang-Mills
  Theory}}, \href{http://dx.doi.org/10.1007/JHEP09(2010)095}{JHEP {\bf 09}
  (2010)  095},
\href{http://arxiv.org/abs/1007.4923}{{\tt arXiv:1007.4923 [hep-th]}}.

\bibitem{Fiol:2020bhf}
B.~Fiol, J.~Mart\'\i{}nez-Montoya, and A.~Rios~Fukelman, \emph{{The planar
  limit of $\mathcal{N}=2$ superconformal field theories}},
  \href{http://dx.doi.org/10.1007/JHEP05(2020)136}{JHEP {\bf 05} (2020)  136},
  \href{http://arxiv.org/abs/2003.02879}{{\tt arXiv:2003.02879 [hep-th]}}.

\bibitem{Passerini:2011fe}
F.~Passerini and K.~Zarembo, \emph{{Wilson Loops in N=2 Super-Yang-Mills from
  Matrix Model}}, \href{http://dx.doi.org/10.1007/JHEP10(2011)065,
  10.1007/JHEP09(2011)102}{JHEP {\bf 09} (2011)  102},
  \href{http://arxiv.org/abs/1106.5763}{{\tt arXiv:1106.5763 [hep-th]}}.
[Erratum: JHEP10,065(2011)].

\bibitem{Fiol:2015mrp}
B.~Fiol, B.~Garolera, and G.~Torrents, \emph{{Probing $ \mathcal{N}=2 $
  superconformal field theories with localization}},
  \href{http://dx.doi.org/10.1007/JHEP01(2016)168}{JHEP {\bf 01} (2016)  168},
\href{http://arxiv.org/abs/1511.00616}{{\tt arXiv:1511.00616 [hep-th]}}.

\bibitem{Bianchi:2019umv}
L.~Bianchi, \emph{{Marginal deformations and defect anomalies}},
  \href{http://dx.doi.org/10.1103/PhysRevD.100.126018}{Phys. Rev. D {\bf 100}
  (2019) no.~12, 126018}, \href{http://arxiv.org/abs/1907.06193}{{\tt
  arXiv:1907.06193 [hep-th]}}.

\bibitem{Beccaria:2020hgy}
M.~Beccaria, M.~Bill\`o, F.~Galvagno, A.~Hasan, and A.~Lerda, \emph{{$
  \mathcal{N} $ = 2 Conformal SYM theories at large $ \mathcal{N} $}},
  \href{http://dx.doi.org/10.1007/JHEP09(2020)116}{JHEP {\bf 09} (2020)  116},
  \href{http://arxiv.org/abs/2007.02840}{{\tt arXiv:2007.02840 [hep-th]}}.

\bibitem{Galvagno:2020cgq}
F.~Galvagno and M.~Preti, \emph{{Chiral correlators in $ \mathcal{N} $ = 2
  superconformal quivers}},
  \href{http://dx.doi.org/10.1007/JHEP05(2021)201}{JHEP {\bf 05} (2021)  201},
  \href{http://arxiv.org/abs/2012.15792}{{\tt arXiv:2012.15792 [hep-th]}}.

\bibitem{Galvagno:2021bbj}
F.~Galvagno and M.~Preti, \emph{{Wilson loop correlators in $\mathcal{N}=2$
  superconformal quivers}}, \href{http://arxiv.org/abs/2105.00257}{{\tt
  arXiv:2105.00257 [hep-th]}}.

\bibitem{Beccaria:2021hvt}
M.~Beccaria, M.~Bill\`o, M.~Frau, A.~Lerda, and A.~Pini, \emph{{Exact results
  in a $ \mathcal{N} $ = 2 superconformal gauge theory at strong coupling}},
  \href{http://dx.doi.org/10.1007/JHEP07(2021)185}{JHEP {\bf 07} (2021)  185},
  \href{http://arxiv.org/abs/2105.15113}{{\tt arXiv:2105.15113 [hep-th]}}.

\bibitem{Beccaria:2021ksw}
M.~Beccaria and A.~A. Tseytlin, \emph{{$1/N$ expansion of circular Wilson loop
  in $\mathcal N=2$ superconformal $SU(N)\times SU(N)$ quiver}},
  \href{http://dx.doi.org/10.1007/JHEP04(2021)265}{JHEP {\bf 04} (2021)  265},
  \href{http://arxiv.org/abs/2102.07696}{{\tt arXiv:2102.07696 [hep-th]}}.

\bibitem{Beccaria:2021vuc}
M.~Beccaria, G.~V. Dunne, and A.~A. Tseytlin, \emph{{BPS Wilson loop in $
  \mathcal{N} $ = 2 superconformal SU(N)
  \textquotedblleft{}orientifold\textquotedblright{} gauge theory and
  weak-strong coupling interpolation}},
  \href{http://dx.doi.org/10.1007/JHEP07(2021)085}{JHEP {\bf 07} (2021)  085},
  \href{http://arxiv.org/abs/2104.12625}{{\tt arXiv:2104.12625 [hep-th]}}.

\bibitem{Billo:2021rdb}
M.~Billo, M.~Frau, F.~Galvagno, A.~Lerda, and A.~Pini, \emph{{Strong-coupling
  results for $ \mathcal{N} $ = 2 superconformal quivers and holography}},
  \href{http://dx.doi.org/10.1007/JHEP10(2021)161}{JHEP {\bf 10} (2021)  161},
  \href{http://arxiv.org/abs/2109.00559}{{\tt arXiv:2109.00559 [hep-th]}}.

\bibitem{Beccaria:2017rbe}
M.~Beccaria, S.~Giombi, and A.~Tseytlin, \emph{{Non-supersymmetric Wilson loop
  in $ \mathcal{N} $ = 4 SYM and defect 1d CFT}},
  \href{http://dx.doi.org/10.1007/JHEP03(2018)131}{JHEP {\bf 03} (2018)  131},
\href{http://arxiv.org/abs/1712.06874}{{\tt arXiv:1712.06874 [hep-th]}}.

\bibitem{Beccaria:2018ocq}
M.~Beccaria and A.~A. Tseytlin, \emph{{On non-supersymmetric generalizations of
  the Wilson-Maldacena loops in $N=4$ SYM}},
\href{http://arxiv.org/abs/1804.02179}{{\tt arXiv:1804.02179 [hep-th]}}.

\bibitem{Cuomo:2021rkm}
G.~Cuomo, Z.~Komargodski, and A.~Raviv-Moshe, \emph{{Renormalization Group
  Flows on Line Defects}}, \href{http://arxiv.org/abs/2108.01117}{{\tt
  arXiv:2108.01117 [hep-th]}}.

\bibitem{Beccaria:2021rmj}
M.~Beccaria, S.~Giombi, and A.~A. Tseytlin, \emph{{Higher order RG flow on the
  Wilson line in $\mathcal{N}=4$ SYM}},
  \href{http://arxiv.org/abs/2110.04212}{{\tt arXiv:2110.04212 [hep-th]}}.

\bibitem{Giombi:2018hsx}
S.~Giombi and S.~Komatsu, \emph{{More Exact Results in the Wilson Loop Defect
  CFT: Bulk-Defect OPE, Nonplanar Corrections and Quantum Spectral Curve}},
  \href{http://dx.doi.org/10.1088/1751-8121/ab046c}{J. Phys. {\bf A52} (2019)
  no.~12, 125401},
\href{http://arxiv.org/abs/1811.02369}{{\tt arXiv:1811.02369 [hep-th]}}.

\bibitem{Giombi:2018qox}
S.~Giombi and S.~Komatsu, \emph{{Exact Correlators on the Wilson Loop in
  $\mathcal{N}=4$ SYM: Localization, Defect CFT, and Integrability}},
  \href{http://dx.doi.org/10.1007/JHEP11(2018)123,
  10.1007/JHEP05(2018)109}{JHEP {\bf 05} (2018)  109},
  \href{http://arxiv.org/abs/1802.05201}{{\tt arXiv:1802.05201 [hep-th]}}.
[Erratum: JHEP11,123(2018)].

\bibitem{Giombi:2021zfb}
S.~Giombi, S.~Komatsu, and B.~Offertaler, \emph{{Large Charges on the Wilson
  Loop in $\mathcal{N}=4$ SYM: Matrix Model and Classical String}},
  \href{http://arxiv.org/abs/2110.13126}{{\tt arXiv:2110.13126 [hep-th]}}.

\bibitem{Nekrasov:2002qd}
N.~Nekrasov, \emph{{Seiberg-Witten prepotential from instanton counting}}, Adv.
  Theor. Math. Phys. {\bf 7} (2004)  831--864,
\href{http://arxiv.org/abs/hep-th/0206161}{{\tt arXiv:hep-th/0206161}}.

\bibitem{Nekrasov:2003rj}
N.~Nekrasov and A.~Okounkov, \emph{{Seiberg-Witten theory and random
  partitions}}, \href{http://dx.doi.org/10.1007/0-8176-4467-9_15}{Prog. Math.
  {\bf 244} (2006)  525--596},
\href{http://arxiv.org/abs/hep-th/0306238}{{\tt arXiv:hep-th/0306238
  [hep-th]}}.

\end{thebibliography}\endgroup

\end{document}